\documentclass[journal,10pt]{IEEEtran}
\usepackage{graphicx} 
\usepackage{array}
\usepackage{comment}
\usepackage{color}
\usepackage{multicol}
\usepackage{multirow}
\usepackage{url} 
\usepackage{amsmath,amssymb,mathrsfs}
\usepackage{algpseudocode,algorithm}
\usepackage{siunitx}
\usepackage{bbm}
\usepackage{booktabs}

\newif\ifconf
\newif\ifskip

\conffalse

\skipfalse

\ifconf
 \else
\fi

\ifCLASSOPTIONcompsoc
  \usepackage[caption=false,font=footnotesize,labelfont=sf,textfont=sf]{subfig}
\else
  \usepackage[caption=false,font=footnotesize]{subfig}
\fi

\newtheorem{prop}{Property}
\newtheorem{cor}{Corollary}
\newtheorem{teo}{Theorem}

\ifCLASSOPTIONcompsoc
  \usepackage[nocompress]{cite}
\else
  \usepackage{cite}
\fi

\newcommand{\example}[2]{
\begin{center}
\parbox{0.9\columnwidth}{
\rule{0.9\columnwidth}{0.5mm}\\
\noindent {\bf Example~#1:}
#2\\
\rule{0.9\columnwidth}{0.5mm}
}\end{center}}

\newcommand{\Fc}{{\cal F}}
\newcommand{\Sc}{{\cal S}}
\newcommand{\Rc}{{\cal R}}
\newcommand{\Kc}{{\cal K}}
\DeclareMathOperator\EE{\mathbb{E}} 
\newcommand{\PP}{\mathbb{P}} 

\definecolor{red}{rgb}{0,0,0}
\definecolor{blue}{rgb}{0,0,0}

\usepackage{soul}

\begin{document}
\begin{sloppypar}

\newcommand{\customsubsection}[1]{\medskip\noindent\textbf{\itshape #1}\medskip}

\title{The RICH Prefetching in Edge Caches \\for In-Order Delivery to Connected Cars}
\author{Ahsan Mahmood$^\dag$, Claudio Casetti$^\dag$, Carla Fabiana Chiasserini$^\dag$, Paolo Giaccone$^\dag$, J\'er\^ome H\"{a}rri$^\ddag$ \\
{\small $^\dag$ Dept.\ Electronics and Telecommunications, Politecnico di Torino, Italy}\\
{\small $^\ddag$EURECOM, 450 route des Chappes, 06410 Sophia Antipolis, France} \\
{\small E-mails:\{ahsan.mahmood,casetti,chiasserini,giaccone\}@polito.it, jerome.haerri@eurecom.fr}
\thanks{A. Mahmood is with Ericsson AB, Sweden. He developed this work as a Ph.D. student at Politecnico di Torino, Italy. C. Casetti, C. F. Chiasserini, and P. Giaccone are with Politecnico di Torino, Italy. They are also with CNIT, and C. F. Chiasserini is a research associate with CNR-IEIIT, Torino, Italy.}
}

\maketitle

\begin{abstract}
Content caching on the edge of 5G networks is an emerging and critical feature to quench the thirst for content of future connected cars. However, the tight packaging of 5G cells, the finite storage capacity at the edge, and the need for content availability while driving motivate the need to develop smart edge caching strategies adapted to the mobility characteristics of connected cars. In this paper, we propose a scheme called RICH (RoadsIde CacHe), which optimally caches content at edge nodes where connected vehicles require it most. In particular, our scheme is designed to ensure in-order delivery of content chunks to end users. {\color{blue} Unlike blind popularity decisions, the probabilistic caching used by RICH accounts for the user mobility information that the system can realistically acquire. Furthermore, we provide a complete system architecture and define the protocols through which  the different system entities can interact. We assess the performance of our  approach against state-of-the-art solutions, under realistic mobility datasets and system scenarios. 
Our RICH edge caching scheme improves significantly the content
availability at the caches and reduces the required backhaul bandwidth,
with beneficial effects for both the end users and the network operators.
} 
\end{abstract}

\IEEEpeerreviewmaketitle

\section{Introduction}\label{sec:intro}

Connected cars are considered by drivers as a projection of their homes on the road, and the same seamless connectivity and content-based services are expected on the road as well.  Communication of connected cars with the network infrastructure can support a wide variety of applications, ranging from critical, safety-related applications to the provision of entertainment services (e.g., video distribution). {\color{blue}An example of the former category is ``See-Through''~\cite{5gaa-seethrough}, an advanced driving assistance automotive use case, where vehicles receive video stream of road conditions from vehicles in front of them to clear the obstructed view and perform actions such as overtaking and lane changing.}
Given the latency-sensitive nature of these applications, a deployment of servers and content at the network edge, hence close to vehicular users, is highly desirable~\cite{priceoffog}. This requirement, along with the expected increase in the number of connected cars,  calls for a significant redesign of the network architecture in order to support high-performance connectivity and reduce the core network congestion due to cloud-based content and applications. 

Mobile Edge Computing (MEC), one of the key technologies for 5G networks, provides computing and storage platforms at the edge of the mobile network~\cite{MEC_etsi,MEC-survey}.  MEC can therefore be used to push content to edge nodes, in close proximity to connected cars. 
A major limitation of this approach, however, is that edge nodes do not have the
same storage flexibility as the cloud, and efficient strategies have to
be developed to store the right content at an Edge Node (EN) (e.g., cellular base station, AP, roadside unit).  
Also, the thirst for wireless capacity
has led to a reduced coverage size of ENs, which requires
content to be replicated in multiple ENs to meet the user demand.

It is worth stressing that, although caching policies have  been widely investigated (e.g.,~\cite{cache_pol3,popularity,cachehit}), 
connected cars add significant challenges to the problem of edge caching. They are, indeed, highly-mobile vehicles, 
thus storage strategies should be optimized {\color{red} not with respect to} the current 
content popularity, rather to the expected content
popularity among users that are about to enter the coverage of ENs. 
Furthermore, the dynamics of connected cars, augmented by the limited
coverage size of ENs, require content to be stored where 
cars have a chance to actually download it, i.e., in slow-speed areas such as congested intersections.

Most of the existing mobility-based caching policies, e.g., in~\cite{edgebuffer,dandapat2013},  require full knowledge of
the trajectory of each car, rising concerns about 
drivers' privacy. Also, they  are not tailored
to in-sequence delivery of content~\cite{cache_pol1, geocaching, cachinghybrid}, typically required by future
on-board streaming applications. A major design challenge for caching
policies  is therefore to rely only on coarse mobility information
(sequences of waypoints, dwell time, etc.), while supporting in-sequence
content delivery.

\begin{figure}[!tb]
\begin{center}
\includegraphics[width=8cm]{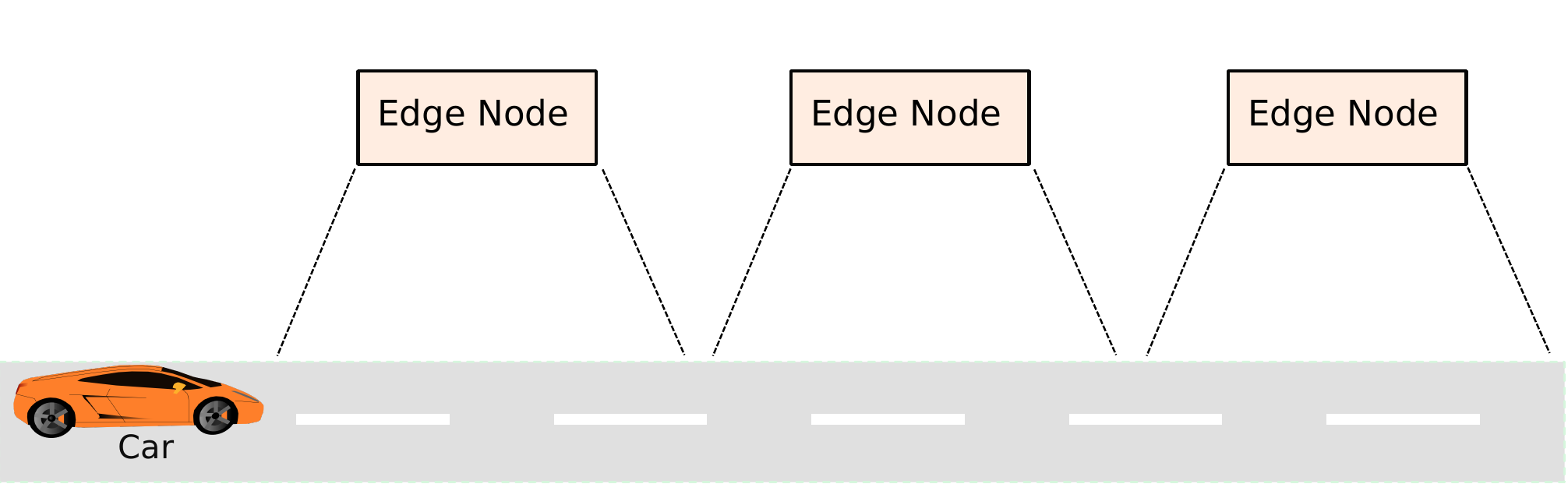}
\caption{Scenario: a car traversing multiple Edge Nodes (ENs).}\label{fig:scenario}
\end{center}
\end{figure}

In this paper, we propose RICH (RoadsIde CacHe), a prefetching policy for  edge caching specifically adapted to
highly dynamic environments with a coarse knowledge of car trajectories. 
We consider an urban environment where cars can connect to ENs, in order to download  {\color{red}content}, as shown in Fig.~\ref{fig:scenario}. 
Our approach is based on the knowledge of the sequence of
  travel waypoints, and some aggregate statistics about the distribution of the dwell time under the coverage of each EN. 
  Note that this is in accordance with the current trend in 5G
systems~\cite{5g}, which foresees a 
dynamic caching system to prefetch content in the ENs, based
on future demand estimation obtained by {\color{red} users' context information,}
such as direction and speed. 

Furthermore, unlike classical works on caching,
 we  consider a data 
streaming application in which the content is divided into fixed-size chunks,   {\color{red}strongly
correlating the download process  at each  EN}, due to the in-sequence chunk delivery. Importantly, we also account for the correlation that mobility introduces   
 in the request process among different ENs.
Indeed, the
instantaneous popularity of a chunk at an EN depends on the actual
temporal and spatial trajectory of all  cars interested in the
corresponding content. The goal of our strategy is thus to
  cache in advance the chunks in the sequence of ENs traversed by the
  car, by choosing the chunks that will be most likely downloaded at
  each specific EN. 
  {\color{red} 
  This  increases  the  {\em cache hit probability}, i.e., the probability that the content chunks are downloaded directly from the cache, which in turn greatly reduces the backhaul traffic and content access delay.}
  
In more detail, our novel contributions are as follows:
\begin{itemize}
\item \emph{Architecture and protocol:} we introduce a split content caching architecture, with a centralized module located in the backhaul, which manages the content stored in the caches located at the ENs. This centralized approach is amenable to being realized through the Software Defined Networking (SDN) paradigm. 
We also define the system protocol governing the interaction between various entities in the vehicular network.
\item \emph{Analytical formulation:} we describe an analytical model capable of predicting the probability of downloading a chunk of a content from a given EN.
\item \emph{Caching scheme:} we leverage the previous model to develop a mobility-aware prefetching strategy that selects what (i.e., which chunk) and where (i.e., in which ENs) to cache, {\color{red}based on the distribution of the dwell times. }
\item \emph{Implementation:} we evaluate our solution under a realistic urban traffic dataset of the city of Bologna. {\color{red} Our simulation model is comprehensive and it closely mimics a real scenario.} We compare our proposed RICH scheme to two state-of-the-art prefetching policies: pure popularity-based caching (POP) and a mobility-aware caching strategy called netPredict~\cite{edgebuffer}.
\end{itemize}


\section{Network Architecture\label{sec:architecture}}

\begin{figure}[!tb]
\begin{center}
\includegraphics[width=8cm]{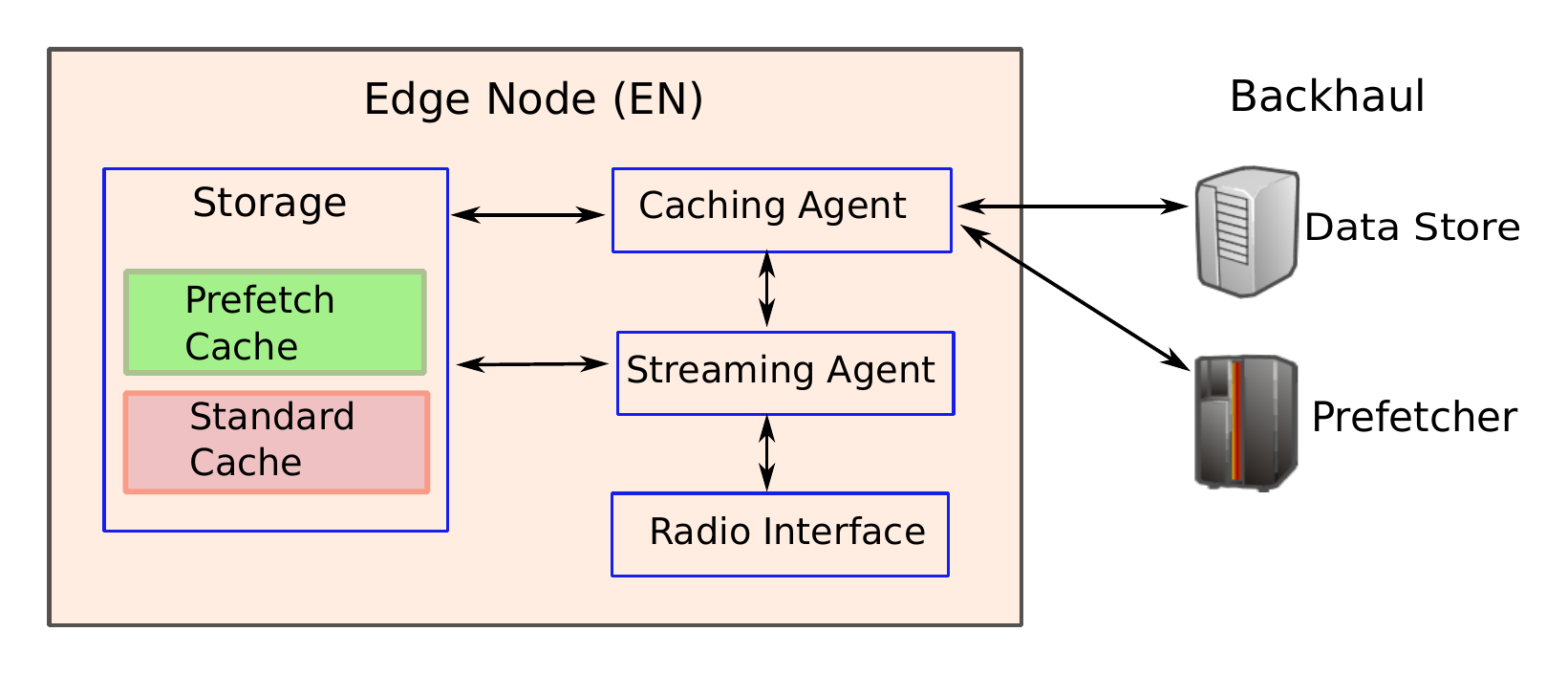}
\caption{Network caching architecture }\label{fig:architecture}
\end{center}
\end{figure}

We consider an urban scenario in which vehicular users traverse coverage areas of several {\color{red}ENs, where each }
EN is responsible for the communication in a specific geographical area, as depicted in Fig.~\ref{fig:scenario}.
Due to the dynamic nature of the vehicular network, along with a large number of users and a volatile wireless channel,  users downloading data  may experience severe service disruptions. Our goal is  to design a caching architecture that provides users with a timely delivery of the content, while also reducing the backhaul bandwidth consumption. 

We focus on a challenging application such as content streaming, for which in-order delivery should be ensured. We let ENs cache parts of the content requested by the vehicular users, and we envision the presence of a {\em Prefetcher} module that instructs, in advance, each EN on which part of the content to store. In this way, the content portion that each user is expected to download from an EN, will be readily available for delivery at the EN when actually reached by the user. 

In the proposed architecture  shown in Fig.~\ref{fig:architecture}, the Prefetcher is located in the network backhaul, along with  a Data Store  module. Both of them are connected with all the ENs. The Data Store provides  a catalog for all the contents, each of which is assumed to be composed of a sequence of {\em chunks}, each univocally identified. The Prefetcher can acquire (e.g., through the user's navigation system) or predict  the sequence of ENs that the user will traverse in the near future.
Also, thanks to past measurements, the distribution of the cars' dwell time under each EN is known to the Prefetcher, as well as  the available storage in the caches of the ENs and of the users' content requests.  Based on such information, the Prefetcher defines {\em which chunks} of {\em which content} should be cached in advance {\em at which EN}, and instructs the ENs accordingly. Upon getting the instructions from the Prefetcher, each EN retrieves the needed chunks from the Data Store before {\color{red}arrival of the user} in the coverage area. This allows leveraging non-real-time data transfer and thus relaxes the QoS requirements for the content transfer into the caches.   

Each EN is responsible for providing streaming service to the user, and, in order to provide timely content delivery, it utilizes a local caching storage, divided in two parts:
\begin{itemize}
\item the {\em Prefetch Cache}, which contains the chunks prefetched from the Data Store, as per Prefetcher's instructions;
\item the {\em Standard Cache}, which stores chunks that were not available in the Prefetch Cache at the time the user requested them, and they have  been retrieved  from the Data Store. 
\end{itemize}
Notably, the latter cache mimics the behavior of a standard caching system, where a new chunk is stored just  after a cache miss is experienced. On the contrary,  the Prefetch Cache behaves similarly to a content delivery network (CDN) node, where the content is proactively stored into the node.

Moreover, the EN carries out its tasks via two agents:
\begin{itemize}
\item the {\em Streaming Agent}  
reads the chunks from the cache and streams them to the user; one separate stream is managed for each user under coverage;
\item the {\em Caching Agent}  
interacts with the Prefetcher, fetches chunks from the Data Store, inserts the {\color{blue}prefetched/missed chunks into the Prefetch/Standard Cache, and manages both the Prefetch and the Standard Cache} 
according to the adopted eviction policy. 
\end{itemize}

\subsection{System protocol}\label{subsec:sysprotocol}

\begin{figure}[!tb]
\begin{center}
\includegraphics[width=8cm]{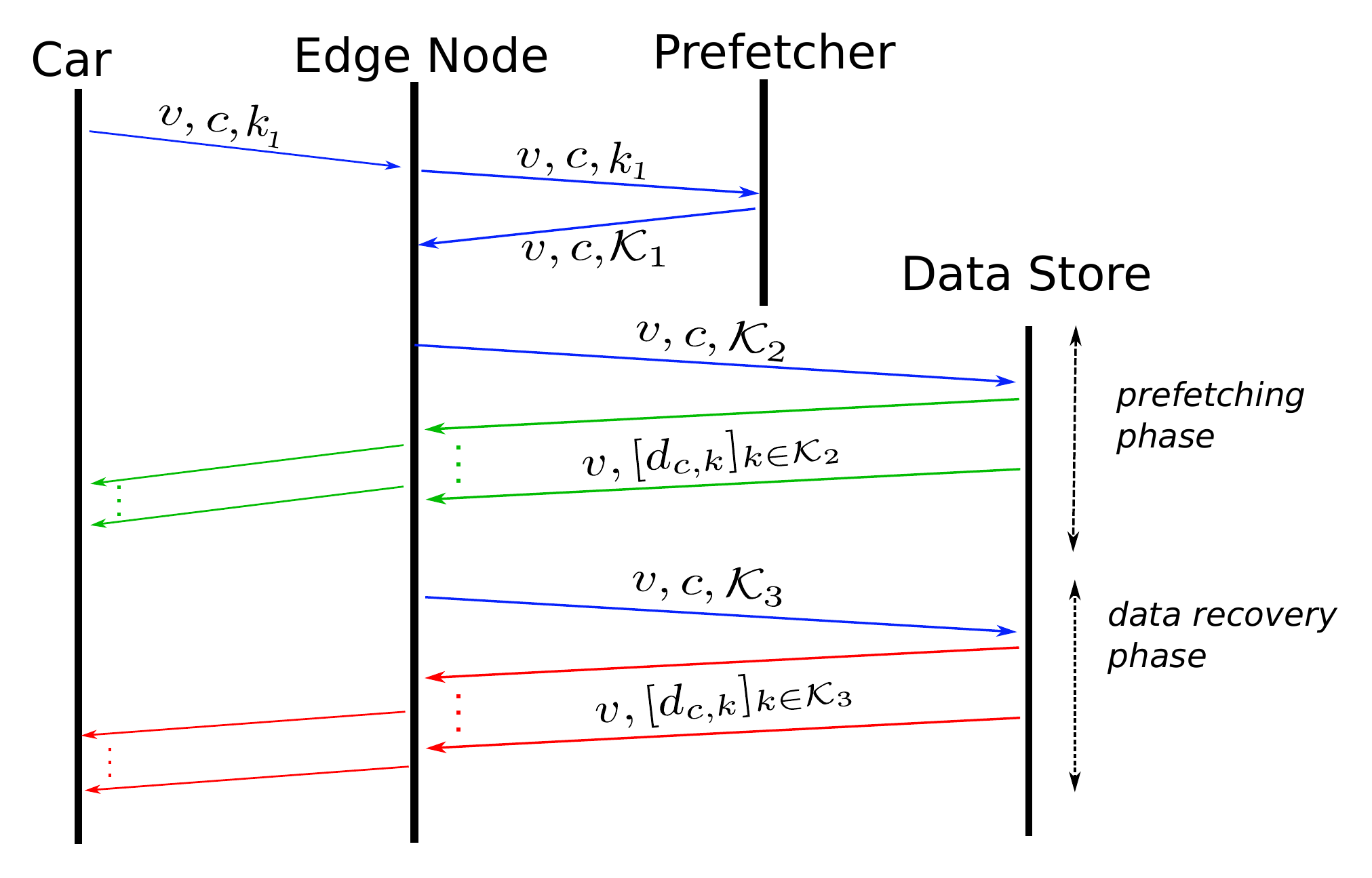}
\caption{Protocol followed by the first EN traversed by a car. }\label{fig:protocol1}
\end{center}
\end{figure}

\begin{figure}[!tb]
\begin{center}
\includegraphics[width=8cm]{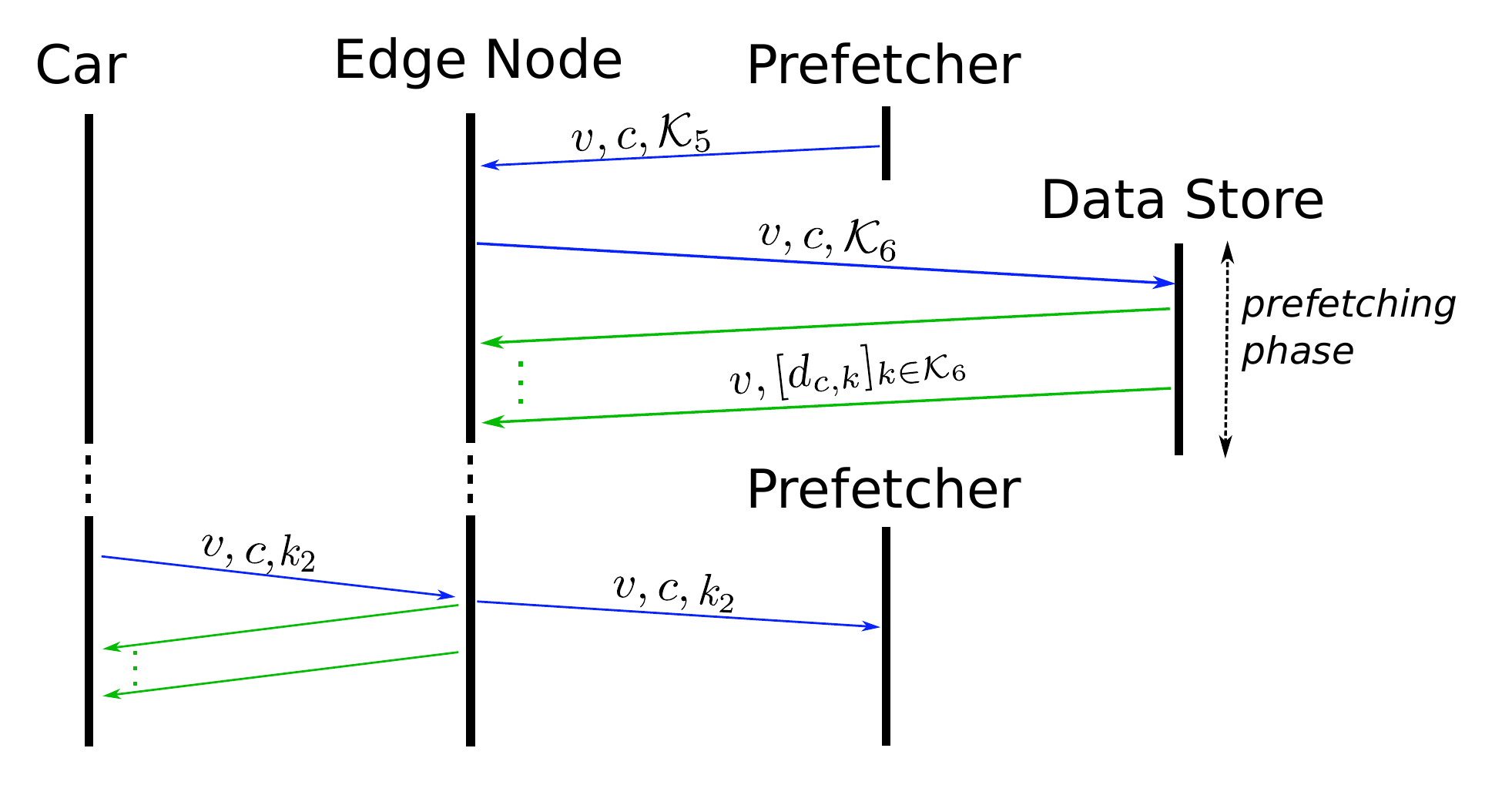}
\caption{Protocol followed by the ENs after the first EN traversed by the car.} \label{fig:protocol2}
\end{center}
\end{figure}

The Streaming Agent contacts the Caching Agent whenever a new content request arrives, as well as whenever a requested chunk is not available in the cache. The Caching Agent, in turn, informs the Streaming Agent as the needed chunk becomes available in the cache. The interactions between the aforementioned entities {\color{red}are} further clarified by the space-time diagrams in Figs.~\ref{fig:protocol1} and~\ref{fig:protocol2}, which depict the protocol followed, respectively,   when a user starts requesting a content and while the car connects with the subsequent ENs to complete the content download, if necessary. In both cases, the proposed protocol is composed of  two phases:
\begin{itemize}
\item \textit{prefetching phase:} an EN prefetches the chunks from the Data Store, as per Prefetcher's instructions, and the EN serves the user on the basis of the chunks stored in the Prefetch Cache; 
\item \textit{data recovery phase:} the EN fetches the missing chunks from the Data Store and sends them in sequence using both the Prefetch and the Standard Cache, if needed.
\end{itemize}
In both figures, chunks downloaded from the Prefetch and from the  Standard Cache are depicted in green and red, respectively.
All control  messages  {\color{red}are} depicted in blue. 

We assume, for simplicity, that a generic content $c$ is divided into equal sized chunks, and we define ${\cal K}_c$ as the set of all the corresponding chunk identifiers. Let $k\in {\cal K}_c$ be a generic chunk identifier of content $c$ and $d_{c,k}$ denote  the corresponding actual data. Let $v$ be a generic user and/or its car {\color{blue} (the notation is summarized in Tab.~\ref{tab:notation}).}

In more details, Fig.~\ref{fig:protocol1} refers to a scenario in which car $v$ enters the coverage area of the first EN along its path and the user requests the first  chunk $k_1$ of content $c$. At the EN, the request is forwarded to the Prefetcher\footnote{The actual implementation of this step is more complex since the request is sent to the Streaming Agent, which registers a new stream for the user and, in turn, forwards the chunk request to the Prefetcher through the Caching Agent. In the following we will omit such level of details in the interaction among the modules internal to the EN.}.
Since the Prefetcher knows the sequence of ENs the car will traverse, it can run the prefetching policy (described in Sec.~\ref{sec:right}) and instructs 
the first EN to prefetch the set $\Kc_1\subseteq\mathcal K_c$ of chunks of content $c$. 
The EN, through the Caching Agent, checks which of the required chunks are already available in the Prefetch Cache and sends a message to the Data Store asking for the missing ones, denoted by $\mathcal K_2 \subseteq\mathcal K_1$. The Data Store sequentially sends the chunks  to the EN, i.e., all  chunks $[d_{c,k}]_{k\in \mathcal K_2}$. The Caching Agent inserts the received chunks in the Prefetch Cache and informs the Streaming Agent about the data availability. Then the Streaming Agent initiates the data stream towards the user, sending the chunks in sequence.

If a chunk is not available in the Prefetch Cache, the system enters the \textit{data recovery phase}, as the data must be retrieved directly from the Data Store.
The latter situation may happen in two cases: (i) some prefetched chunks were evicted from the cache to make space for other chunks; (ii) all the prefetched chunks have been transmitted and the car is  still under coverage. In both cases, the Streaming Agent attempts to serve the car by getting the chunks from the Standard Cache. Otherwise, the missing chunks, denoted by $\mathcal K_3\subseteq \mathcal K_c$, are fetched from the Data Store for transmission to the user.  
{\color{red} $\mathcal K_3$ is chosen large enough to compensate for the round-trip time from the Prefetch Cache to the Data Store and to guarantee a continuous streaming from the Data Store to the user, using the EN as relay until the car leaves the coverage area.}

Fig.~\ref{fig:protocol2} shows only the prefetching phase occurring at the second EN and at the subsequent ENs traversed by the car. The eventual data recovery phase is omitted as it is identical to the one in Fig.~\ref{fig:protocol1}. When the car enters the coverage of the first EN, the Prefetcher instructs all the subsequent ENs to prefetch the required chunks before the arrival of the car in their coverage areas. {\color{blue}This ensures that the user experiences no delay due to content prefetching in the subsequent ENs\footnote{\color{blue} Note that the delay due to non-overlapping coverage areas, if any, can be avoided by assuming a sufficiently large playout buffer at the user; the buffer size would depend on the type of network deployment.}.} As shown in Fig.~\ref{fig:protocol2}, the EN receives a message to prefetch the chunks $\mathcal K_5\subseteq \mathcal K_c$. For the subset $\mathcal K_6$ of chunks in $\mathcal K_5$ that are not yet available in the cache, the EN asks the Data Store and stores the corresponding chunks in the Prefetch Cache.

When the car enters the EN coverage, it sends a chunk request for the first missing chunk $k_2$. 
The EN forwards this message to the Prefetcher in order to notify it about the car entrance in the coverage area. The EN is responsible to send all the required chunks to the car which are available in the local cache. Eventually, the Prefetcher may trigger a data recovery phase to compensate for missing chunks in the cache.


\begin{table}[!tb]
\centering
\caption{{\color{blue}Notation}\label{tab:notation}}
\begin{tabular} {cl}
\toprule
\textbf{$v$}& A generic user and/or its car \\
\textbf{$c$}& Content identifier \\
\textbf{$k$}& Generic chunk identifier \\
\textbf{$\mathcal K_c$}& The set of all chunk identifiers of content $c$ \\ 
\textbf{$d_{c,k}$}& Data corresponding to chunk $k$ of content $c$ \\ 
\multirow{2}{*}{\textbf{$\mathcal K_i$}}& A generic subset of $\mathcal K_c$ representing the chunks \\ & of content $c$ instructed-to-be/actually prefetched \\ 

\bottomrule
\end{tabular}
\end{table}

\section{RICH prefetching policy\label{sec:right}}

{\color{blue} Given the caching architecture discussed above, our goal is to maximize the Prefetch Cache hit probability across all cars. Indeed,  the user  throughput increases whenever the desired chunks are locally downloaded from the EN and not from the backhaul, while leading to better utilization of the backhaul network resources. We therefore propose an efficient strategy to be run at the Prefetcher, which  selects the chunks to store in the Prefetch Cache (hereinafter simply called cache) of the ENs traversed by a car. We recall that the car path is expressed as the sequence of traversed ENs, and the distribution of the dwell time under the ENs is obtained through historical data.

Note that the above caching problem could  be formalized as an ILP problem  by discretizing the time into steps and associating a  binary decision variable with {\em each} step, {\em each} EN, and {\em each} chunk of {\em every content}, expressing whether that chunk of that particular content is stored in a particular EN at that time step. Then, assuming that the sequence of ENs traversed by  the cars, the time at which they enter an EN, and the cars content demand are known, we could express formally the hit probability to maximize over all cars and all chunks. However, such formulation would need global knowledge of the system dynamics over an extended period of time; also, given that the cars dwell time is random, its solution would require stochastic optimization methods.  In light of the limited knowledge that is available in practice and of the problem complexity,  below we design an approximated method, which, as shown in Sec.\,\ref{sec:results}, proves to be very efficient.}

We start by motivating our approach using a toy-case example and highlighting the benefits of knowing the distribution of the cars dwell time under an EN.  Then we theoretically evaluate the probability that a specific chunk is downloaded from an EN by a tagged car, which will be used to define our prefetching policy. Indeed, our
approach consists in letting each EN store those chunks whose 
probability to be downloaded by a car is above a given
threshold.

 
\subsection{A toy-case example}


Consider a single EN covering the area of an intersection controlled by a traffic light. The dwell time of each car and, thus, the number of downloaded chunks depends on the traffic light state. Let us now focus on the subset of cars that request a specific content starting from its first chunk. Assume that 80\% of these cars experience green light, thus the corresponding dwell time under the EN is small and the users (labeled by ``fast'') can download up to 10 chunks.
The 20\% of these cars experience red, thus the corresponding dwell time under the EN is large and the users (labeled by ``slow'') can download up to 100 chunks. Hence, the average number of chunks that can be downloaded by a car is 28.

Assume now that the system adopts a prefetching policy that makes its decisions based on the average number of chunks downloaded by the users, as, e.g., the state-of-the-art netPredict policy discussed later in Sec.~\ref{sec:netpre}.
Such a policy will store just the first 28 chunks of the content in the cache.
This can be seen as an inefficient decision for  both kinds of users. Indeed, for the fast users, able  to download just the first 10 chunks, the additional 18 chunks stored in the cache are completely useless and a waste in terms of cache occupancy.  For the slow users, able to download up to 100 chunks, the cache is heavily underutilized since only the first 28 chunks are available.

In our work we propose instead to exploit the distribution of the dwell time under each EN to increase the performance of the overall caching system. Indeed, in the above example, {\color{blue} two more clever solutions can be easily envisaged. The first is to store 10 chunks only: the slow users still experience cache misses as in the netPredict case, however we now avoid the waste of cache occupancy in the case of fast users and save room for other content. 
The second solution, instead, 
 stores the first 100 chunks: in this way, all the users will find the requested chunks in the cache, at the cost of a higher cache occupancy.}

Note that the possible inefficiencies arising in a single EN (i.e., cache) scenario are exacerbated in the presence of a sequence of ENs, due to the required in-sequence delivery of the chunks.
Thus, we advocate the use of prefetching policies that are aware of the dwell-time distribution, and not only of the average dwell time under an EN, as, e.g., netPredict does.

\subsection{Chunk download probability for large caches}\label{subsec:chunkdownloadprob}

We now derive the probability that a user successfully downloads a chunk from the cache. 
Let us focus on one car and one specific  content that a user wishes to download from the traversed ENs. Assume that EN $i$, with $1\leq i\leq N$, is the $i$th EN along the path of the car, composed of $N$ ENs.

We start by defining the chunk delivery process under a best-case scenario in which all ENs cache the whole content. This implicit assumption of
very large cache is aimed at devising a simple model that will be
tailored to small caches later in this section. 

Let $Y_i$ be the random variable representing the last chunk received 
from EN $i\geq 1$,  and let $Y_0=0$ by definition. 
Then the set of chunks downloaded from EN $i$ is given by: 
\(
\{k | k\in(Y_{i-1},Y_i]\}
\).
Thus the probability that chunk $k$ is downloaded from EN $i\geq 1$ is, for any $k\in\{1,\ldots,K\}$, 
\begin{equation}\label{eq:1}
\phi_i(k)\mathord{=} \PP \left (k\in(Y_{i-1},Y_i]\right) \mathord{=} \PP
  \left (Y_i\geq k ~\land~ Y_{i-1}<k \right)
\end{equation}
Note that if $N$ is large enough, the content will be downloaded for sure from some EN, thus $\sum_{i=1}^\infty \phi_i(k)=1$. 

Let $X_i$ be the random variable representing the total number of chunks
downloaded from EN $i$ by the user. The probability density
function (pdf) of $X_i$ depends mainly on two factors: (1) the mobility
of the car, since, e.g., longer dwell times under the EN coverage
typically imply larger amounts of download data, and (2) the actual
throughput obtained by the user when connected to the EN,
which in  turn depends on the wireless data rate and on channel
contention among other users. In Sec.~\ref{subsec:scenario}, we  will describe how to compute
the pdf of $X_i$ in the urban scenario under study. 

Based on our definitions,  it is easy to see that for $i\geq 1$,
\begin{equation}\label{eq:3}
Y_i=Y_{i-1}+X_i=\sum_{j=1}^i X_j \,.
\end{equation}
The theorem below relates the download probability $\phi_i(k)$  to $X_i$. 
\begin{teo}
Given a car traversing a sequence of ENs, the probability to download a specific chunk $k$ from  EN $i$, with $1\leq i\leq N$, can be expressed as
\begin{equation}\label{eq:phi}
\phi_i(k)=\sum_{n=1}^{k-1} \PP \left (X_i\geq k-n | Y_{i-1}=n\right)\PP \left (Y_{i-1}=n \right) \,.
\end{equation}
\label{teo1}
\end{teo}

\begin{IEEEproof}
Given (\ref{eq:3}),  we can write \eqref{eq:1}, for any $i\geq 1$, as:
\begin{eqnarray}
\phi_i(k) &=& \PP \left (Y_{i-1}+X_i\geq k ~\land~ Y_{i-1}<k \right)\nonumber\\
&=& \sum_{n=1}^{k-1} \PP \left (X_i\geq k-Y_{i-1}|Y_{i-1}=n \right)
\PP \left (Y_{i-1}=n \right) \nonumber\\
&=& \sum_{n=1}^{k-1} \PP \left (X_i\geq k-n |Y_{i-1}=n\right)\PP \left (Y_{i-1}=n \right) \,. \nonumber
\label{eq:phi-proof}
\end{eqnarray}
\end{IEEEproof}
Note that, as expected, for $i=1$ the expression in \eqref{eq:phi} becomes
\(
\phi_1(k)=\PP(X_1\geq k)
\)
since the user will download chunk $k$ from the first EN only if
the total amount of downloaded chunks from EN $1$ is greater than $k$.
Now, thanks to the well-known property of the expectation of non-negative integer random variables, we can claim:
\begin{prop}
$\sum_{k=1}^{K}\phi_i(k)=\EE[X_i] \,.$
\end{prop}

The following corollary holds in the special case when the car dwell times under the ENs are
i.i.d. random variables. 
Note that this case is addressed here for completeness, as well as to provide a more explicit expression of the $\phi$'s, however
our approach does not require such an assumption.
\begin{cor}\label{cor:1}
Let the   random variables $X_i$'s be i.i.d. and defined on a positive
support. Let $f_{X}(k)$ be their discrete pdf. Then, for any $i\geq
1$, we have:
\begin{equation}\label{eq:phi_conv}
\phi_i(k)=\left (f_X \ast \phi_{i-1}\right )(k)
\end{equation}
where $\ast$ is the convolution operator. 
\end{cor}
The proof is reported in Appendix~\ref{sec:app}.
{\color{red} The complexity of computing the convolution for all ENs is $O(N^2 K^2)$. Notably, such computation can be done offline, based on past statistics.}

We now consider the  case of an EN cache of limited size.
Let $M_i$ be the available space in terms of chunks at EN $i$.
Then we can introduce a discrete random variable,
$\widehat{X}_i$, which represents the total number of cached chunks 
downloaded from EN $i$ by the user,  given the available room in
the cache of EN $i$. The pdf of $\widehat{X}_i$ is given by:
\begin{equation}\label{eq:finite}
\PP\left (\widehat{X}_i= x  | M_i \right ) = \left \{ 
\begin{array}{lc}
\PP (X_i =x) &  0\leq x < M_i\\
\PP (X_i \geq x)& x = M_i\\
0 &  x > M_i \,.
\end{array}
\right . 
\end{equation}
Indeed, it is not possible to download more than $M_i$
cached chunks, and the events corresponding to a number of downloaded chunks 
larger than $M_i$ in the original model with large cache size, now correspond
to downloading all the $M_i$ available chunks.  

Given the above distribution, the probabilities $\phi_i(k)$ can be
computed as in (\ref{eq:phi}), or when $\widehat{X}_i$'s
are i.i.d.\ as in (\ref{eq:iid-conv}) .

To further clarify the behavior of the download probability at each EN, we provide  an example below.

\example{1}{
Consider a toy scenario in which 
a car traverses four ENs and large-sized caches are available.
$X_i$'s  are i.i.d.\ with a symmetric triangular distribution and mean value equal to 10 chunks, i.e., 
the average number of chunks downloaded at each EN is 10.
Fig.~\ref{fig:phi} shows the download probabilities $\phi_i(k)$ at each EN $i$ for each chunk $k$, computed by applying  Corollary~\ref{cor:1}.
From Fig.~\ref{fig:phi}, we observe that at the first EN $\phi_1(k)$  decreases as $k$ increases 
since the randomness 
in the mobility reduces the probability of downloading further chunks. Due to the limited support of the 
distribution of $X_1$, $\phi_1(k)$ becomes zero for $k\geq 20$ chunks. 
At the second EN, $\phi_2(k)$ is now bell-shaped, since values of $k$ close to zero correspond to the case 
in which $X_1$ takes very small values (which is unlikely), i.e., the car speed is very high under 
the coverage of the first EN, hence 
the user does  not have enough time to download any chunk. The maximum is obtained around 15, which is reasonable 
since in the case of deterministic mobility with $X_i=10$ chunks for any $i$, the chunks to be downloaded 
would be exactly in the interval $[10,20]$, which is symmetric around 15.  
The chunk download probability from the following ENs ($i>2$) still exhibits a bell-shaped behavior, 
but with a larger support. This is 
due to the higher uncertainty on downloading a specific chunk from a given EN, which, in turn, is due to the increased 
randomness  in the number of previously delivered chunks.
}


\subsection{The RICH prefetching algorithm\label{subsec:cache}}

Based on the previous definition of the chunk download probability, we define our prefetching policy. 
The goal of our scheme is to  ensure that the probability with which
a user can download a chunk from any of the ENs is greater than a given threshold
$\tau$, with $\tau\in [0,1]$.  Our scheme thus identifies
the {\em smallest} set of ENs that should cache each chunk so that its download probability exceeds 
 $\tau$.  This set  depends on the combined probabilities of downloading the chunk from each EN.
 Thus, setting $\tau$ allows  controlling the maximum amount of chunks to be downloaded  
from the backhaul. 

The pseudocode of RICH is reported in Alg.\,\ref{algo:th}, which, for each
chunk $k$, returns the set of ENs, $\Sc(k)$, that must store $k$ according to the RICH prefetch policy.
After the initialization, for each chunk $k$ (lines~\ref{ln2}-\ref{ln3}),  RICH considers the set $\mathcal F(k)$ of corresponding download probabilities (line~\ref{ln1}). To avoid degenerate solutions, RICH just considers the ENs for which it is possible to download chunk $k$, i.e.\ $\phi_i(k)>0$. 
Now the algorithm iterates (lines~\ref{ln4}-\ref{ln5}) on all the download probabilities in decreasing order (line~\ref{ln6}).
Chunk $k$ is now stored at the EN $i$ corresponding to the highest $\phi_i(k)$ value. If $\phi_i(k)$ is
already greater than $\tau$, no other EN should cache $k$. Otherwise,
the EN corresponding to the second top value of $\phi_i(k)$ should
store the chunk too. Eventually, chunk $k$ will be cached at as
many  ENs,  associated to the top $\phi_i(k)$ values, as necessary so
that the sum of their $\phi_i(k)$ exceeds $\tau$. 
Notably, the probabilities can sum up since referring to disjoint events. 

\begin{algorithm}[!tb]
\caption{RICH prefetch with a single-threshold}
\label{algo:th}
\begin{algorithmic}[1]
\Require $\tau$ \Comment{Probability threshold}
\Require $\{\phi_i(k)\}_{i,k}$\Comment{Download probability for any chunk and for any EN}
\For {$k=1,\ldots,K$}\Comment{For any chunk $k$}\label{ln2}
\State {$\Sc(k)=\emptyset$, $p_k=0$}\Comment{Init}
\State {$\Fc(k) \leftarrow \{\phi_i(k)|\phi_i(k)>0\}_{i}$} \Comment{Store download prob.\ {\color{red} in descending order} for chunk $k$ }\label{ln1}
\While{$\Fc(k) \neq \emptyset$ and $p_k\leq \tau$}\Comment{Check threshold}\label{ln4}
\State {$p_{top} \, \leftarrow \,$ remove the highest prob.\ from $\Fc(k)$}\label{ln6}
\State {$i_{top} \, \leftarrow \,$ EN index corresponding to $p_{top}$}
\State {$p_k=p_k+p_{top}$} \Comment{Accumulate the probabilities} 
\State {$\Sc(k) =\Sc(k)\cup \{i_{top}\}$}\Comment{Update the list of ENs} \label{ln7}
\EndWhile\label{ln5}
\EndFor\label{ln3}
\State \Return {$\{\Sc(k)\}_k$}\Comment{ENs where to prefetch each chunk  }
\State \Return $p_k$ \Comment{Final download probability}
\end{algorithmic}
\end{algorithm}

At the end of the procedure, $p_k$ represents the estimated download 
probability of chunk $k$ from {\em any} EN in $\Sc(k)$, thus with $|\{\Sc(k)\}|$ copies of chunk $k$.
Whenever $p_k\geq \tau$, the prefetching procedure has run successfully. Otherwise, it fails since it is not possible to find any set of {\color{red}ENs} where to prefect the chunk in order to satisfy $\tau$. In such a case, we do not store any copy of the chunks. Note also that the last ENs of the path experience a significant {\em border effect} since there are not subsequent ENs where to prefetch the chunks. 
{\color{red}In Alg.~\ref{algo:th}, the computation complexity of each loop is \( O(N\log N+N)\), due to the sorting in  line~\ref{ln1} and the fact that each chunk can be stored in at most $N$ ENs. Thus the overall complexity is $O(KN\log N)$.
}

To understand the effect of setting $\tau$, we consider some extreme cases. 
If we set $\tau=0$, a single copy of each chunk is stored in the  most probable EN.
Whereas, if we set $\tau=1$ and the threshold is reached, the chunk is stored in any EN for which the download is possible. 
Thus the value of $\tau$ can be numerically optimized  to maximize a given performance metric, i.e., the overall {\color{red} cache hit} probability.

\begin{figure}[!tb]
\begin{center}
\includegraphics[width=7cm]{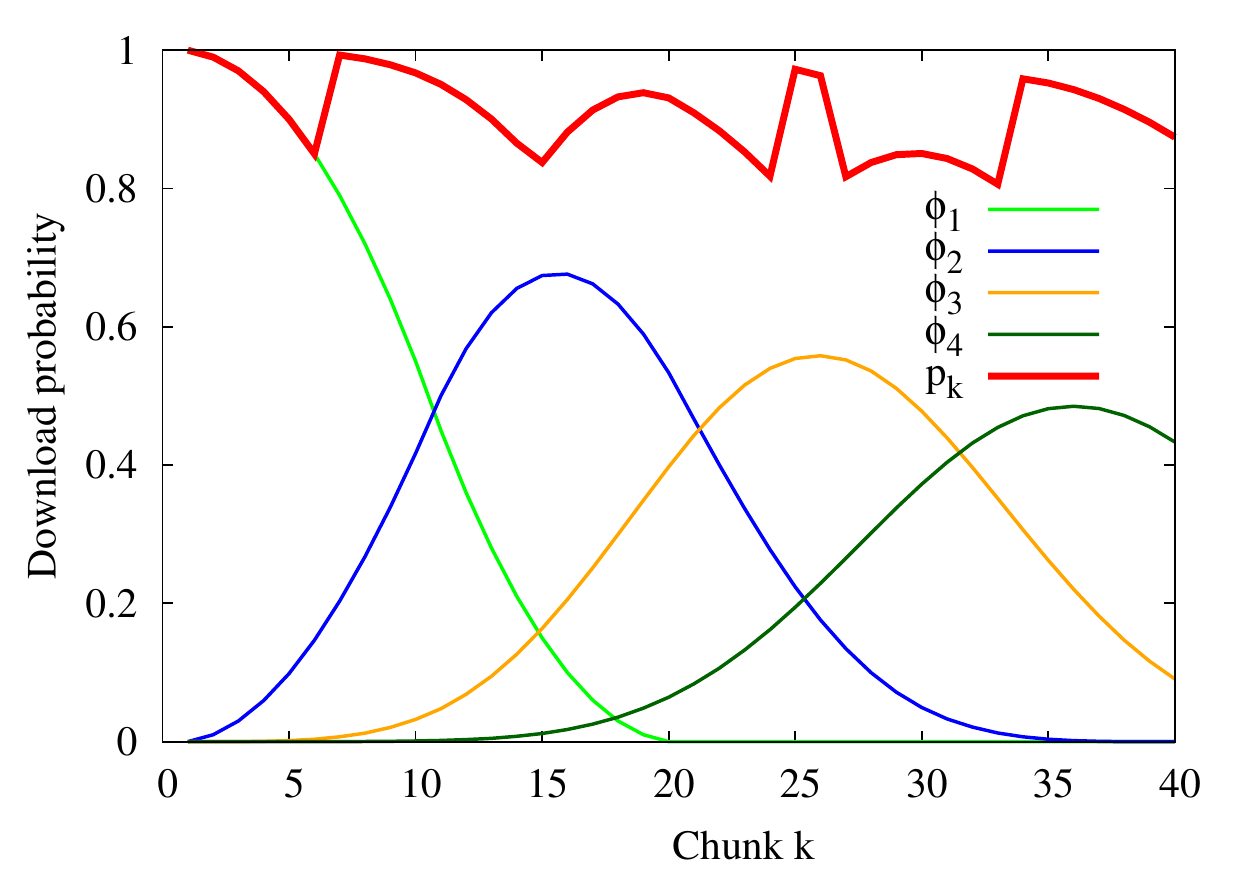}
\caption{Download probability for each EN and overall download probability, $p_k$, 
for the RICH caching policy with a single threshold $\tau=0.8$, given a toy example scenario with  $\EE[X_i]=10$ chunks.}\label{fig:phi}
\end{center}
\end{figure}

\begin{figure}[!tb]
\begin{center}
\includegraphics[width=7cm]{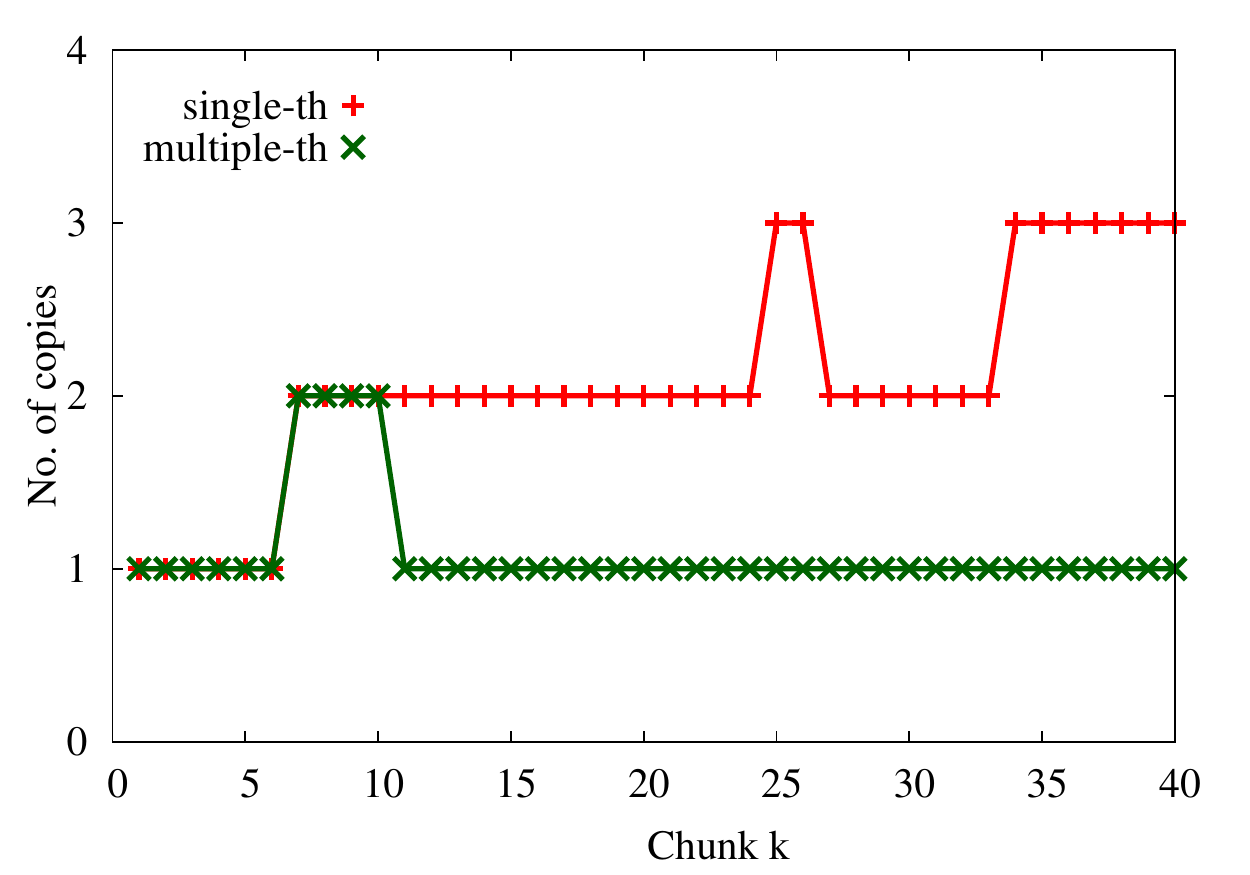}
\caption{Number of copies for the single-threshold and multiple-thresholds RICH prefetching scheme.}
\label{fig:phi_copies}
\end{center}
\end{figure}

\begin{figure}[!tb]
\begin{center}
\includegraphics[width=7cm]{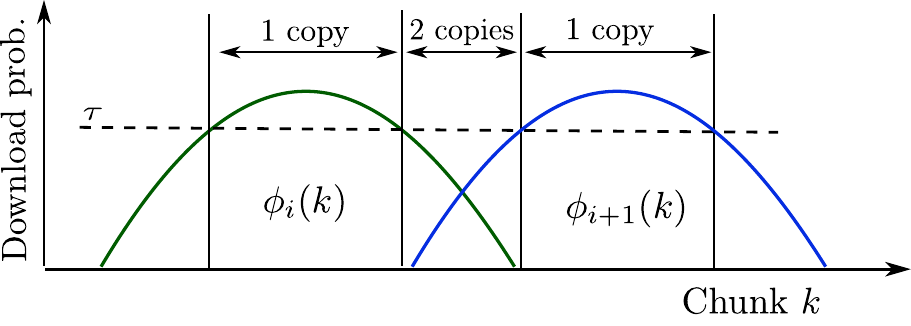}
\caption{Example of number of copies for two subsequent ENs.}
\label{fig:toy-rsu}
\end{center}
\end{figure}

\begin{figure}[!tb]
\begin{center}
\includegraphics[width=7cm]{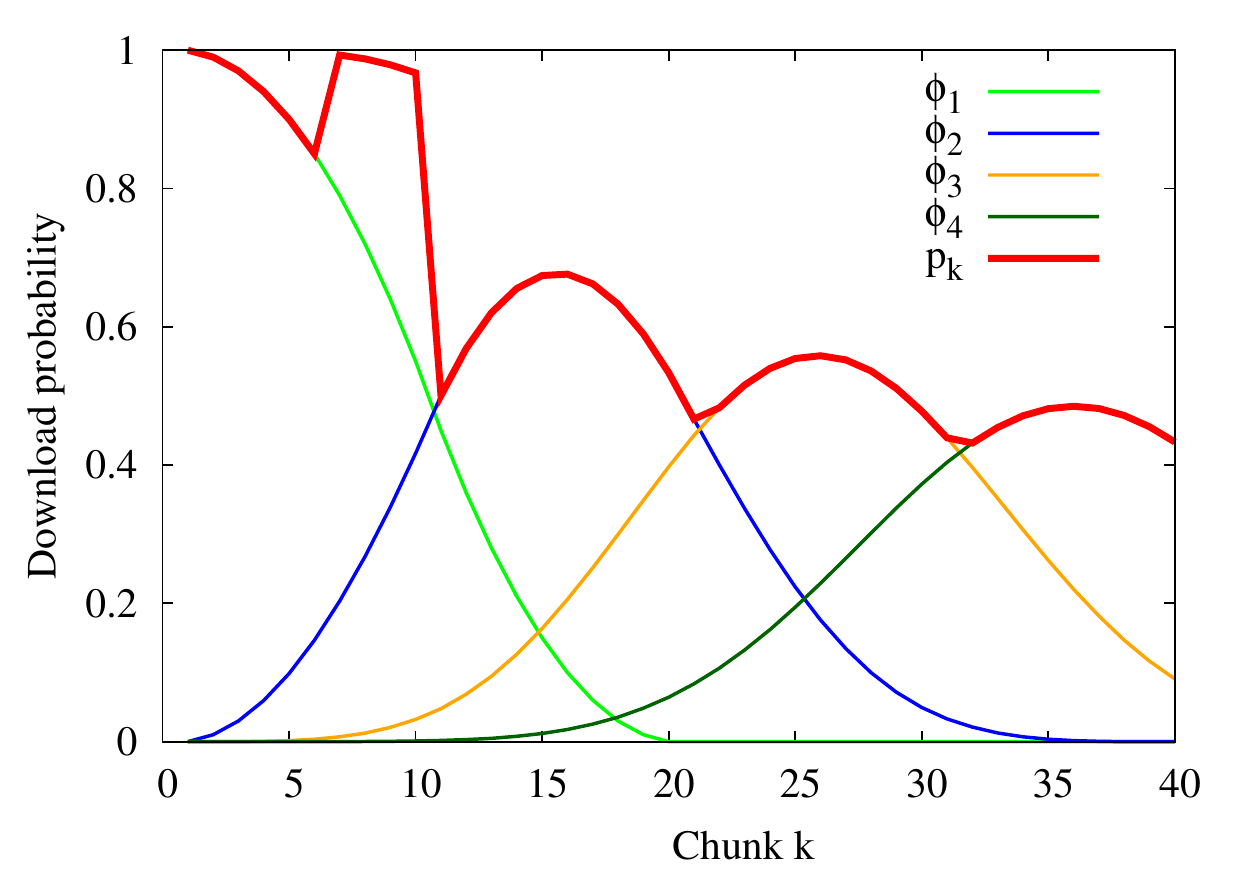}
\caption{Download probability for each EN and overall download probability, $p_k$, 
for the RICH caching policy with multiple thresholds, being $\tau_1=0.8$ (for $k\in [1,10]$) and $\tau_2=\tau_3=\tau_4=0.4$ (for $k>10$).}
\label{fig:phi_multiple}
\end{center}
\end{figure}

In the example of Fig.~\ref{fig:phi},
we show the different $\phi_i(k)$ corresponding to a simple triangular distribution for
$X_i$ with $E[X_i]=10$ chunks. We also show the final $p_k$ obtained with RICH by setting $\tau=0.8$. 
The non-monotonic behavior of $p_k$ is due to the different
number of copies that are stored at the ENs for different chunks, as shown by the curve labeled ``single-th" in Fig.~\ref{fig:phi_copies}. The behavior of the number of copies is due to two different effects.
Intuitively, as $k$ increases,
the uncertainty about the possibility to download chunk $k$ increases, 
thus the RICH caching algorithm compensates it by creating  a higher number of
copies. Indeed, in Fig.~\ref{fig:phi_copies} the number of copies grows from 1 to 3. At the same time,  the uncertainty is usually larger for the chunks which are ``between'' two ENs and thus the number of copies can be non-monotonic, as shown around chunk 26 in Fig.~\ref{fig:phi_copies}. To understand this second effect, consider the example shown in Fig.~\ref{fig:toy-rsu}, depicting $\phi_i(k)$ and $\phi_{i+1}(k)$ for two subsequent ENs. The most probable chunks at each EN (whose download probability is greater than $\tau$) will be stored as just one copy, whereas the other chunks (between the most probable chunks) will be stored as two copies, due to the uncertainty in the precise moment when the car will enter  EN $i+1$.

\subsubsection{Coping with uncertainty}

We now propose two further enhancements to our scheme, in order to efficiently cope with chunk requests that are farther in the future.

{\bf Multi-threshold RICH policy.}
Our intuition is that a fixed threshold $\tau$, as the one used in Alg.\,\ref{algo:th}, may be unfit for  chunks that are expected to be downloaded in farther ENs, for which the level of uncertainty is larger. Indeed, for such chunks the single-threshold RICH policy will naturally require a very large number of copies, thus wasting precious space in the cache that could be better used for other users.  We can therefore consider  a dedicated threshold $\tau_i$ for each EN $i$, leading to a decreasing sequence of thresholds $\{\tau_i\}_i$. 
The actual values of the thresholds can be optimized numerically to maximize the {\color{red} cache hit} probability.
We define instead the range of chunks for which threshold $\tau_i$ is adopted for EN $i$ as the set of all the chunks such that $\phi_i(k)\geq \phi_j(k)$, for any other EN $j\neq i$.  The reason for this choice is that the chunks with the largest download probability at one EN must affect the actual threshold value to use at such EN. Importantly, the complexity of optimizing the thresholds grows as $N$, i.e., the number of ENs in the path, and not as the number of chunks in a content (which may be arbitrary large, depending on the specific kind of content).

Fig.~\ref{fig:phi_multiple} shows the same scenario as in Fig.~\ref{fig:phi} but using different values of threshold for the four ENs. Only $\tau_1=0.8$, while all the other thresholds are equal to $0.4$. The final $p_k$ is lower with respect to the single threshold case for $k>10$, but this is due to the smaller number of stored copies (usually 1 for most of the chunks), as shown in Fig.~\ref{fig:phi_copies}. This reflects the intuition that for the chunks whose uncertainty is large, it is better not to ``waste'' cache storage with multiple copies.

{\bf Refreshing the policy.} 
Due to the high level of uncertainty, for {\color{red}``far-in-the-future''} chunks, RICH  compensates with a high number of copies, wasting a large amount of cache storage while achieving a marginal performance improvement. Thus, we {\color{red}design} RICH to prefetch contents just on the initial sequence of ENs, and to refresh the prefetching decision before entering the first EN not considered in the previous decision. This approach permits to better trade the effectiveness of the prefetching decision with the time to prefetch the content.

\subsubsection{Eviction policy}

Consider  now a generic scenario with multiple users served by one EN. Since the EN cache is finite, we need an eviction policy  determining which chunks should be
removed when the cache is full and which new chunks must be
inserted. We adopt the following policy: the chunks removed with higher priority are the ones that have been already delivered, i.e., those destined to users not anymore under coverage. Among these chunks,  the ones with the lowest download probability are evicted first. In this way, we can efficiently exploit the cache under chunk demands that vary in both  space and  time.


\section{Simulation scenario and methodology\label{sec:scenario_methodology}}
Here,  we first introduce 
the scenario and the real-world vehicular traces used  to investigate the performance of RICH. Then we describe the  methodology adopted for the comparison of  RICH against  state-of-the-art solutions.

\subsection{Reference scenario}\label{subsec:scenario}
 
\begin{figure}[tb!]
	\centering
		\includegraphics[width=0.6\columnwidth]{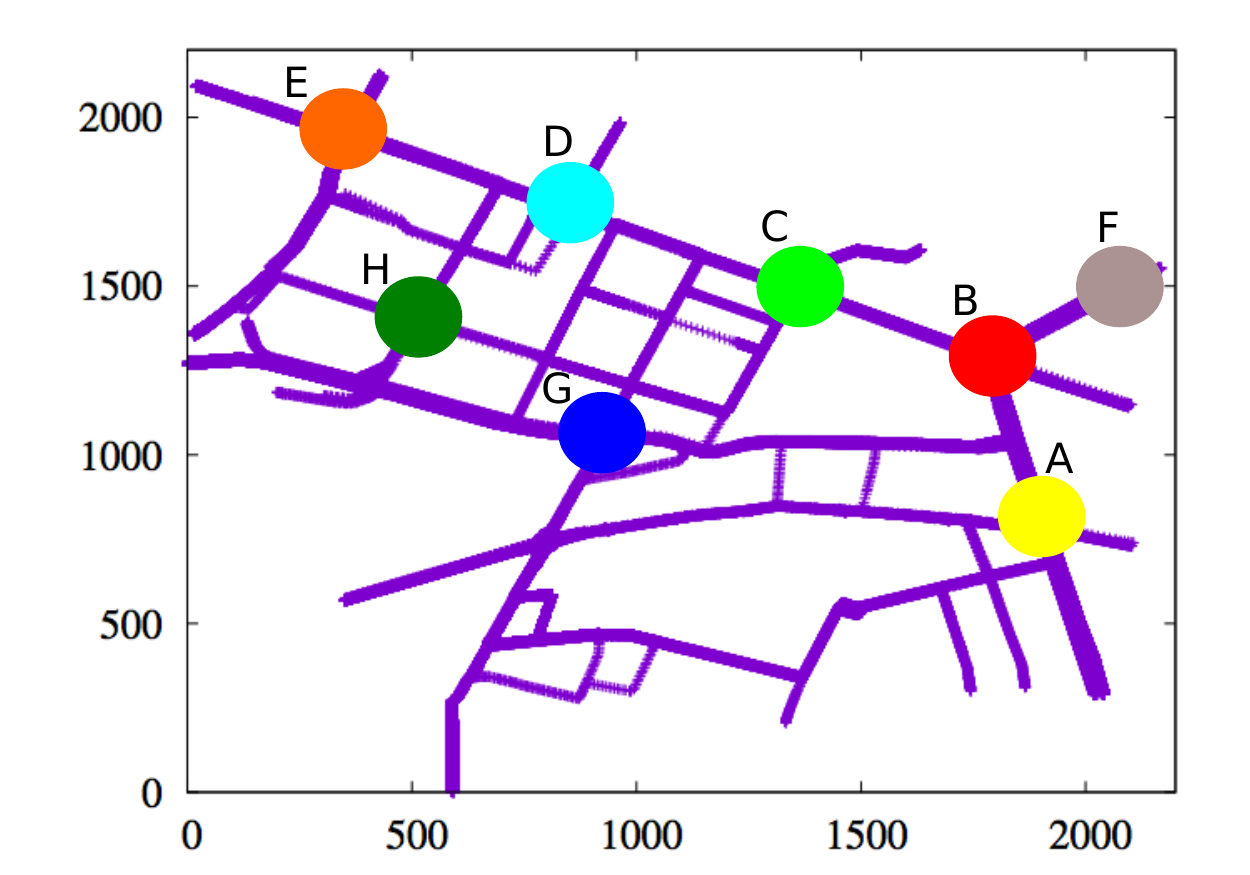}
	\caption{Reference urban area in the city of Bologna. Circles represent EN locations and the corresponding coverage areas (distance is expressed in meters).}
	\label{fig:PasubioJoint}
\end{figure}

We take as reference scenario 
a real-world 2~km $\times$ 2~km urban area of the Italian city of Bologna, illustrated in 
Fig.~\ref{fig:PasubioJoint}. The detailed vehicular mobility traces 
were obtained to reproduce the experimental data gathered through the traffic detectors available at the intersections~\cite{bologna2004}.  
The total trace duration is 79 minutes 
comprising 11,079 vehicles (approximately, 950 are
simultaneously on the map) and representing 120 minutes of the morning rush hours, under quite stationary traffic conditions.

In the above urban section, we select the major traffic arteries and place eight ENs along them, as represented in Fig.~\ref{fig:PasubioJoint} in correspondence of main intersections regulated by a traffic light.  Let $\Rc$=\{A,B,C,D,E,F,G,H\} be the set of  ENs;  
all ENs have the same cache size. We assume an ideal radio range of 100~m at each EN, thus {\color{blue} the coverage areas of  subsequent ENs (denoted by circles in the figure) do not overlap.   Note however that, even if we consider here non-overlapping coverage areas, our scheme can be applied under any scenario.}

To investigate the effect of the prefetching policy across multiple ENs,  
we consider {\color{red} only those cars in the trace}  that enter the coverage of any three ENs. Among the large number of the possible combinations of  three ENs, we consider only the ones with a large number of users and call them  {\em significant paths}.
In total, we identify 7 significant paths with a minimum number of cars equal to 45 each (see Tab.~\ref{tab:paths}  for details), obtaining 2,053 cars in total  following a significant path. 
 Given the vehicular traces of all the cars traversing the significant paths, for each EN we derive the empirical distribution of the car dwell times, needed by the Prefetcher.

Finally, although RICH  can be implemented on any Infrastructure-to-Vehicle technology (ITS-G5/DSRC, WiFi or LTE),  we opted for the IEEE 802.11a WiFi standard, yet operating on the 5.9 GHz frequency band. The reasons for this choice are twofold. 
First, since our application is not strictly safety related, we would not use ITS-G5 to transmit on the ITS-G5 bands (at least 5.875-5.905 GHz). Nevertheless, we want to avoid transmitting on the WiFi bands at 5.4\,GHz so as  to not interfere with non-vehicular communications and increase the  capacity available for content streaming delivery. Second, ETSI is preparing a deregulation of the ITS band to allow WiFi access, e.g., 802.11a, under a `detect-and-avoid' principle against ITS-G5 technology~\cite{etsi-103319,EURECOM5191}. Thus, according to C-ITS~\cite{c-its}, we test  RICH on the Service Channel 1 (SCH 1) at 5.875-5.885 GHz and, accordingly, ignore potential coexistence with ITS-G5.

\subsection{Simulation methodology}\label{subsec:methodology}

We developed a discrete-event simulator, 
based on OMNeT++~\cite{omnet}, which models the network architecture in Fig.~\ref{fig:architecture} with 8 deployed ENs, as  in Fig.~\ref{fig:PasubioJoint}. 
Even if in our model we implemented both the Prefetch and the Standard Caches, in the following we will investigate just the performance due to the Prefetch Cache.

We assume an ideal communication link from the ENs to the Prefetcher, with a propagation delay equal to $10~\mu$s, corresponding to a Prefetcher located physically in the same urban area.
Instead, the Data Store is connected to the ENs, with a 100~Mbps communication link and a  propagation delay of 2~ms.
We model the  wireless access network with a detailed 802.11a model provided in Inet-Framework 3.3~\cite{inet}. The simulation parameters are defined in Table~\ref{tab:sim_par}.

\begin{table}[tb!]
\centering
\caption{OMNeT++ simulation parameters}
\label{tab:sim_par}
\resizebox{\linewidth}{!}{%
\begin{tabular}{l	l	l	l}
\toprule
\multicolumn{1}{l}{\textbf{Parameter}} & \multicolumn{1}{l}{\textbf{Value}} & \multicolumn{1}{l}{\textbf{Parameter}} & \multicolumn{1}{l}{\textbf{Value}} \\ \midrule
WiFi                  & 802.11a                 & WiFi active scan               &     false   \\
Frequency band     & 5 GHz                   &    Beacon interval             &     100 ms  \\
Bandwidth             &  10 MHz                 &  Receiver sensitivity          &     -85 dBm \\
Max bitrate           &  54 Mbps                &   SNIR threshold               &     4 dB   \\
Rate control          & AARF                    &  Background noise power        &     -101 dBm \\
MTU                   & 1 kbytes                &  Pathloss type                 &     Free space \\
Tx power              & 33 dBm                  &   Pathloss coefficient         &     3   \\
Antenna gain          & 1 dB                    &   Max communication range      &     110 m   \\
Retry limit           & 7                       &     Max interference range     &     200 m   \\ \bottomrule       \end{tabular}
} 
\end{table}

Each significant path comprises 3 ENs: we run the prefetching policy and evaluate the performance at the first two, to avoid the border effect due to the last EN, described in Sec.~\ref{subsec:cache}. 
Table~\ref{tab:no-users} shows the statistics about the temporal average of the number of users under each EN (given that at least one car is present), and the total number of distinct cars under coverage.
EN A is located at the most crowded intersection, thus we expect heavy congestion and lower per-user throughput. On the other hand, H is at the least crowded intersection, for which we expect the least number of users on average, thus high per-user throughput.

\begin{table}[!tb]
\centering
\caption{Number of cars under each EN in Bologna urban area\label{tab:no-users}}
\resizebox{\linewidth}{!}{%
\begin{tabular} {ccccccccc}
\toprule
\textbf{EN}& A & B & C & D& E &F &G &H \\ \midrule
\textbf{Avg. cars}& 27.51 & 13.36 & 7.88 & 4.52 & 8.35& 6.02& 3.42& 2.79 \\
\textbf{Total cars}& 1580& 1510& 527& 382& 337& 1456& 161& 206 \\ \bottomrule
\end{tabular}
}
\end{table}

\begin{table}[!tb]
\centering
\caption{Significant paths identified in Bologna urban area\label{tab:paths}}
\resizebox{\linewidth}{!}{%
\begin{tabular}{cccccccc}
\toprule
\textbf{Path}& ABC& ABF& CDE& FBA& FBC& HDC& HGA \\ \midrule
\textbf{No. of cars}& 54& 555& 337& 810& 91& 45& 161\\ \bottomrule
\end{tabular}%
}
\end{table}

To evaluate the download probability for any chunk and for any EN, i.e.\ $\{\phi_i(k)\}_{i,k}$, we must determine the number of downloaded chunks $X_i$ at each EN.
Let $W_i$ be the random variable of the dwell time of a generic car under EN $i$. Let $b$ be the total bandwidth available when a user is under coverage of an EN. Let $u_i$ be the average number of cars under the coverage of EN $i$. Let $s$ be the total size in bits of each chunk. Approximatively, we can claim that:
\begin{equation}\label{eq:x}
X_i=\dfrac{W_i \cdot b}{s\cdot u_i}. 
\end{equation}
Replacing the empirical values of $W_i$ and $u_i$ (given in Table~\ref{tab:no-users}) in \eqref{eq:x}, we estimate the distribution of $X_i$, and then the empirical $\phi_i(k)$ for any EN $i$ and chunk $k$.
Finally, we evaluate $\widehat{X}_i$ to take into account  the finite cache, by applying \eqref{eq:finite}. 

{\color{blue}We remark that, we account for content popularity while evaluating the performance of RICH. Indeed, each} time a new car enters 
the coverage of any EN for the first time, it generates a content request according to a Zipf's distribution with exponent $\alpha=0.75$, 
coherently with the value observed in \cite{lfu}.
The size of one content is 
2,600 chunks, with each chunk being  65~kbytes large. 
 The normalized cache size is defined as the cache size
  divided by the total size of all contents in the catalog. The size of the catalog is 10 contents. 

The performance metrics we consider are as follows\footnote{The above statistics are computed considering the total time  period during which a car was under the coverage of any of the ENs.}: 
\begin{itemize}
\item {\em cache throughput} $[{\rm bit/s}]$: average amount of data received by the user over time and directly downloaded from the cache;
\item {\em cache hit probability}, $P_{hit}\in [0,1]$: fraction of chunk requests that are satisfied by directly downloading the chunk from the cache, i.e., in the event of a cache hit;  
\item {\em backhaul traffic} $[{\rm bit/s}]$: average amount of data downloaded from the server  over time, i.e., due to the event of a cache miss;
\item {\em normalized backhaul overhead}: total amount of additional traffic transferred between the Data Store and the ENs with respect to amount of traffic delivered to the users, normalized to the traffic delivered to the users; it can be also negative in case of {\em content reuse} in the cache;
\item {\em normalized cache size}, $\hat{C}\in [0,1]$: size (i.e.,  maximum allowed occupancy) of each cache, normalized to the catalog size;
\item {\em network cache occupancy} $\in [0,N]$: average total cache occupancy across the ENs, normalized to the catalog size.
\end{itemize}

\subsection{Alternative prefetching policies}\label{sec:netpre}

Below we describe our benchmark schemes.

{\bf POP:} it is a mobility-agnostic approach that prefetches contents in decreasing order of popularity till storage saturation. 
It requires  knowledge of the popularity level of all the contents in advance, but it is provably optimal in terms of hit probability for a single cache and under a stationary content request process~\cite{lfu}.
We tailor the behavior of POP to our data streaming scenario, in which each content is divided into chunks. POP  stores the chunks of the most popular contents in sequence; if the cache space is not sufficient for the whole content, 
only the initial chunks of the content are stored.

{\bf netPredict~\cite{edgebuffer}}: it 
exploits  both  spatial and temporal predictions of the car path based on the previous history. 
For a fair comparison with RICH, we assume that the spatial prediction in netPredict is perfect and, hence, the sequence of ENs traversed by the car is known in advance.
Based on the knowledge of the {\em average} dwell time under each EN and the value of the average bandwidth available at each EN, netPredict stores a number of chunks given by their product. Thus, the adopted communication model in~\cite{edgebuffer} is identical to~\eqref{eq:x}, but exploits just $E[X_i]$ instead of the distribution of $X_i$ as in RICH. In a nutshell, netPredict can be seen as a special case of RICH for deterministic $X_i$.

\subsection{Statistical analysis of the reference scenario}

Cars may arrive under the coverage of the ENs following various paths, as shown in Table~\ref{tab:paths}. Also, the number of cars on each path is different. Considering such traffic conditions, the dwell time distributions (and hence the distributions of $\widehat{X}_i$) of the cars at the ENs show interesting statistical properties (see the values in Table~\ref{tab:statdesc} for a cache size of 2,600 chunks). We observe that,  if incoming cars at some EN $i$ take different paths (with a significant number of cars in each path), the distribution of $\widehat{X}_i$ shows high value of skewness and kurtosis. For example, Table~\ref{tab:paths} shows that EN B is involved in several paths. Since cars enter B's coverage with different incoming and outgoing roads, the distribution of $\widehat{X}_B$ shows the highest value of skewness and kurtosis. A similar behavior can be observed in case of ENs A, D, and H. On the other hand, the incoming cars under the coverage of E and G follow only the paths CDE and HGA, respectively. Hence, the distributions of $\widehat{X}_E$ and $\widehat{X}_G$  show lower values of skewness and kurtosis. 

\begin{table}[!tb]
\centering
\caption{Statistical description of empirical $\widehat{X}_i$ in Bologna urban area.\label{tab:statdesc}}
\resizebox{\linewidth}{!}{%
\begin{tabular} {ccccccccc}
\toprule
\textbf{EN} & A & B & C & D& E &F &G &H \\ \midrule
\textbf{Skewness}& 2.11 & 2.48 & 0.41 & 1.74 & -0.11 & -0.15 & -0.18 & 1.62 \\
\textbf{Kurtosis}& 8.57 & 11.60 & 2.71 & 6.89 & 2.63 & 2.22 & 1.68 & 5.98 \\ \bottomrule
\end{tabular}
}
\end{table}


\section{Numerical results} \label{sec:results}


As the first step, we use an exhaustive approach to find the combination of the thresholds $\tau_i$ at the three ENs for maximizing the {\color{red} cache hit} probability. Table~\ref{tab:opt_th} shows the values we obtained for different cache sizes, assumed to be the same at all ENs. In general, the optimal threshold value  in the first EN, $\tau_1$, is higher than the others due to the smaller uncertainty on the cars  mobility in the first hop. 
By construction, $\tau_3$ depends on the chunks that are most probably downloaded from EN 3. However, such chunks can also be downloaded from EN 2. The higher value of $\tau_3$ as compared to $\tau_2$ allows some additional chunks to be stored in EN 2, which increases the {\color{red} cache hit} probability. For cache size $\geq 7800$ chunks, the optimal value of the thresholds does not change, because the large cache capacity is never fully utilized.

\begin{table}[!tb]
\centering
\caption{Optimal thresholds for different cache sizes.}
\label{tab:opt_th}
\begin{tabular}{ccccc}
\toprule
\multicolumn{2}{c}{\textsc{Cache size}} & \multicolumn{3}{c}{\textsc{Optimal threshold}} \\ \cmidrule(lr){1-2} \cmidrule(lr){3-5}
\#contents & \#chunks & $\tau_1$ & $\tau_2$ & $\tau_3$ \\
\midrule
1& 2600 & 0.88&0.67&0.70 \\
2&5200 & 0.92&0.58&0.67 \\
3&7800 & 0.99&0.57&0.67 \\
4&10400 & 0.99&0.57&0.67 \\
5&11000 & 0.99&0.57&0.67 \\
\bottomrule
\end{tabular}
\end{table}

\begin{figure*}[tb!]
	\centering
		\includegraphics[width=0.31\textwidth]{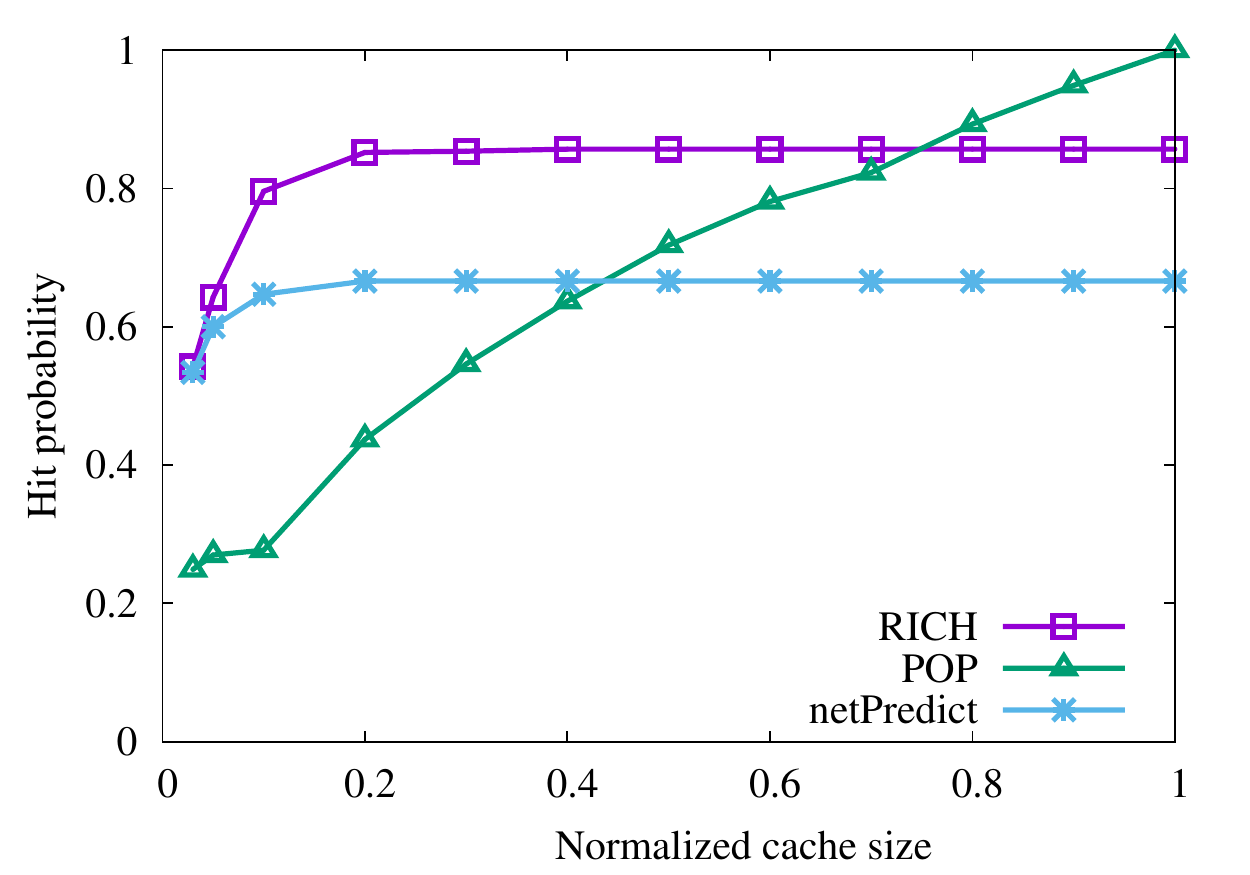}
		\includegraphics[width=0.31\textwidth]{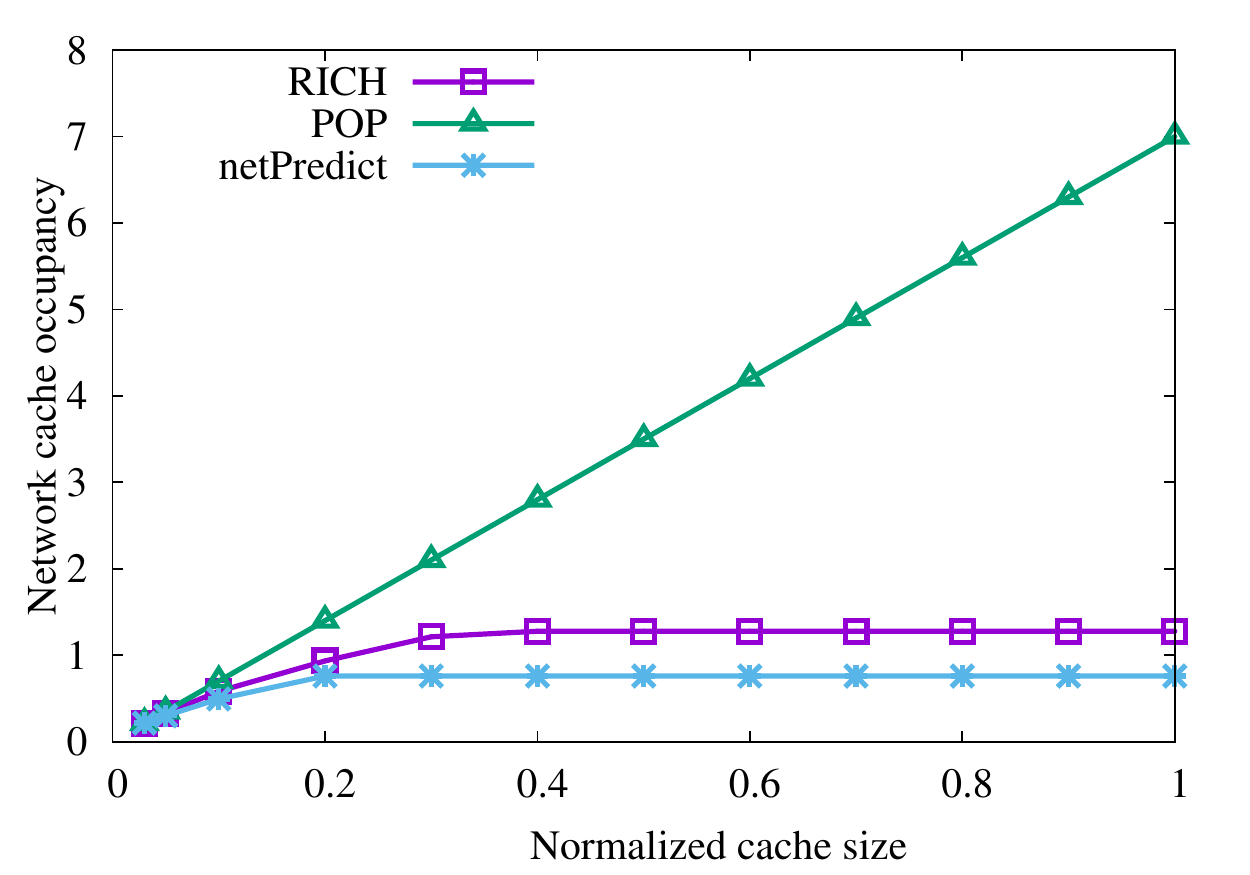}
		\includegraphics[width=0.31\textwidth]{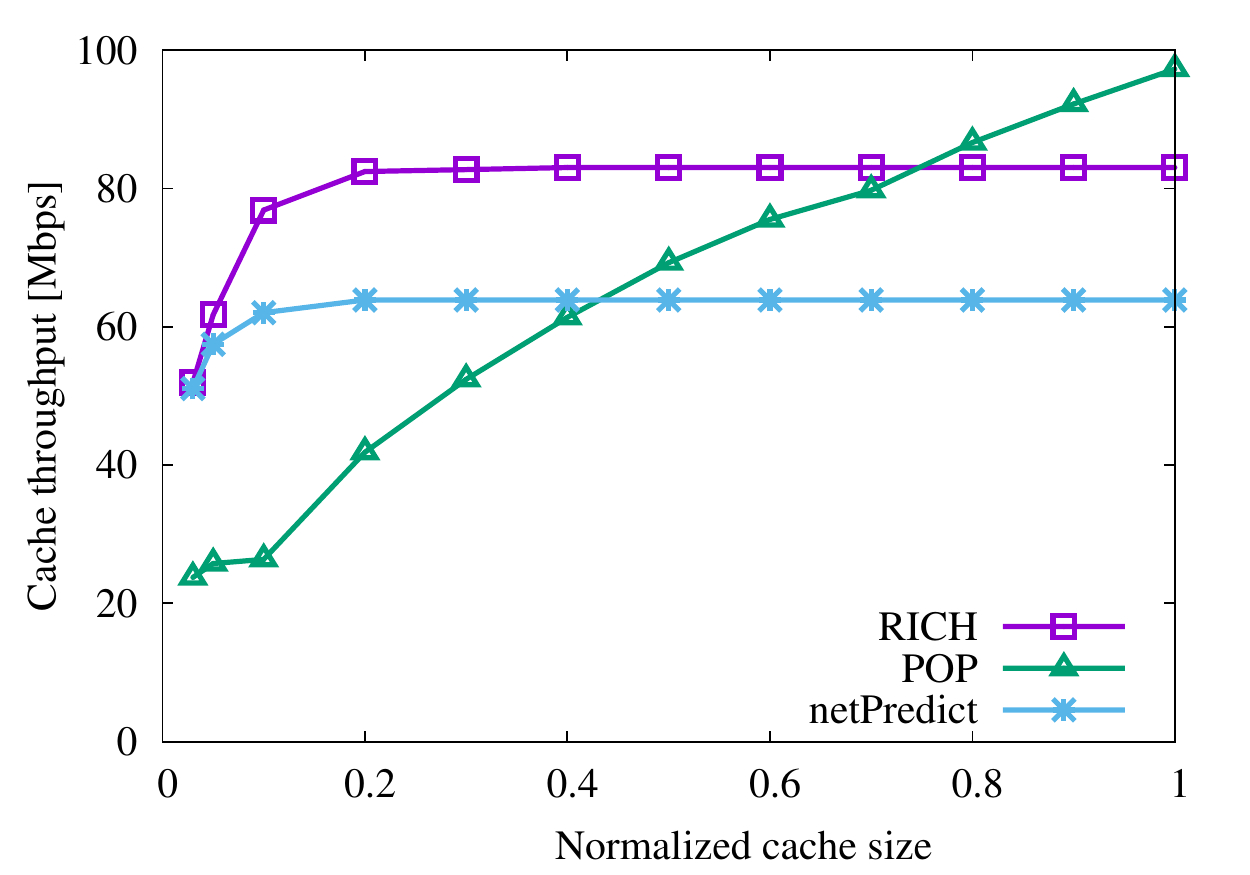}
	\caption{Urban Bologna scenario: cache hit probability (left); network cache occupancy (center); cache throughput (right).}
	\label{fig:hit}
\end{figure*}
\begin{figure*}[tb!]
	\centering
		\includegraphics[width=0.31\textwidth]{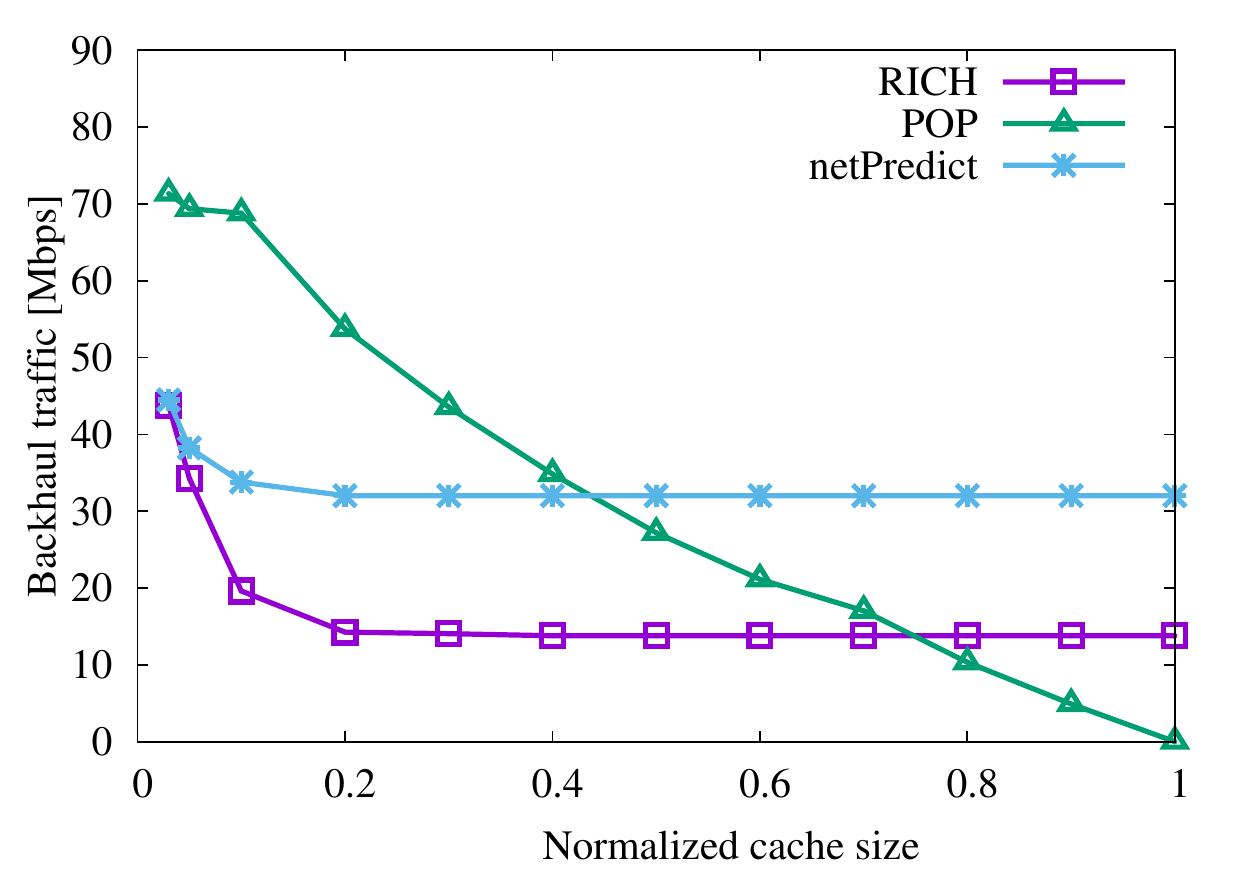}
		\includegraphics[width=0.31\textwidth]{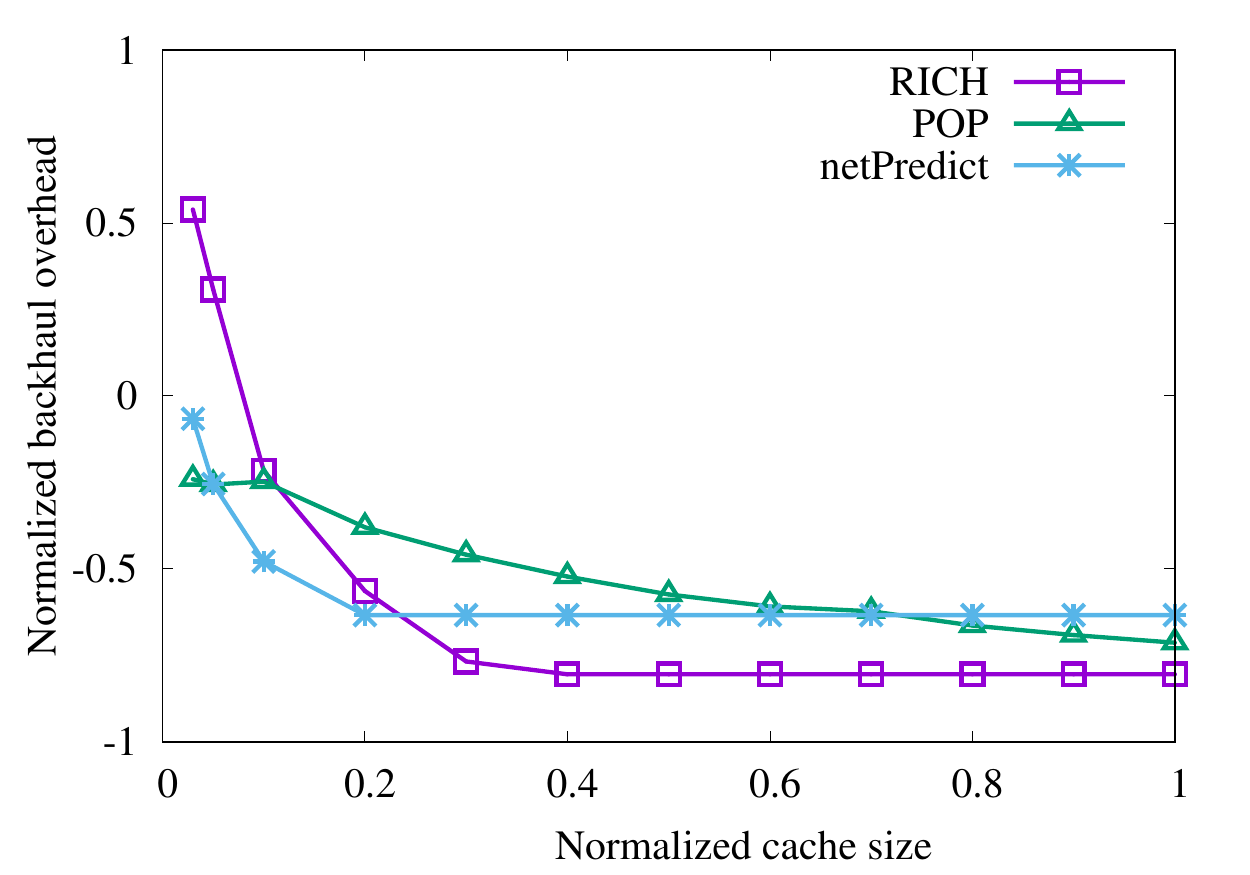}
	\includegraphics[width=0.31\textwidth]{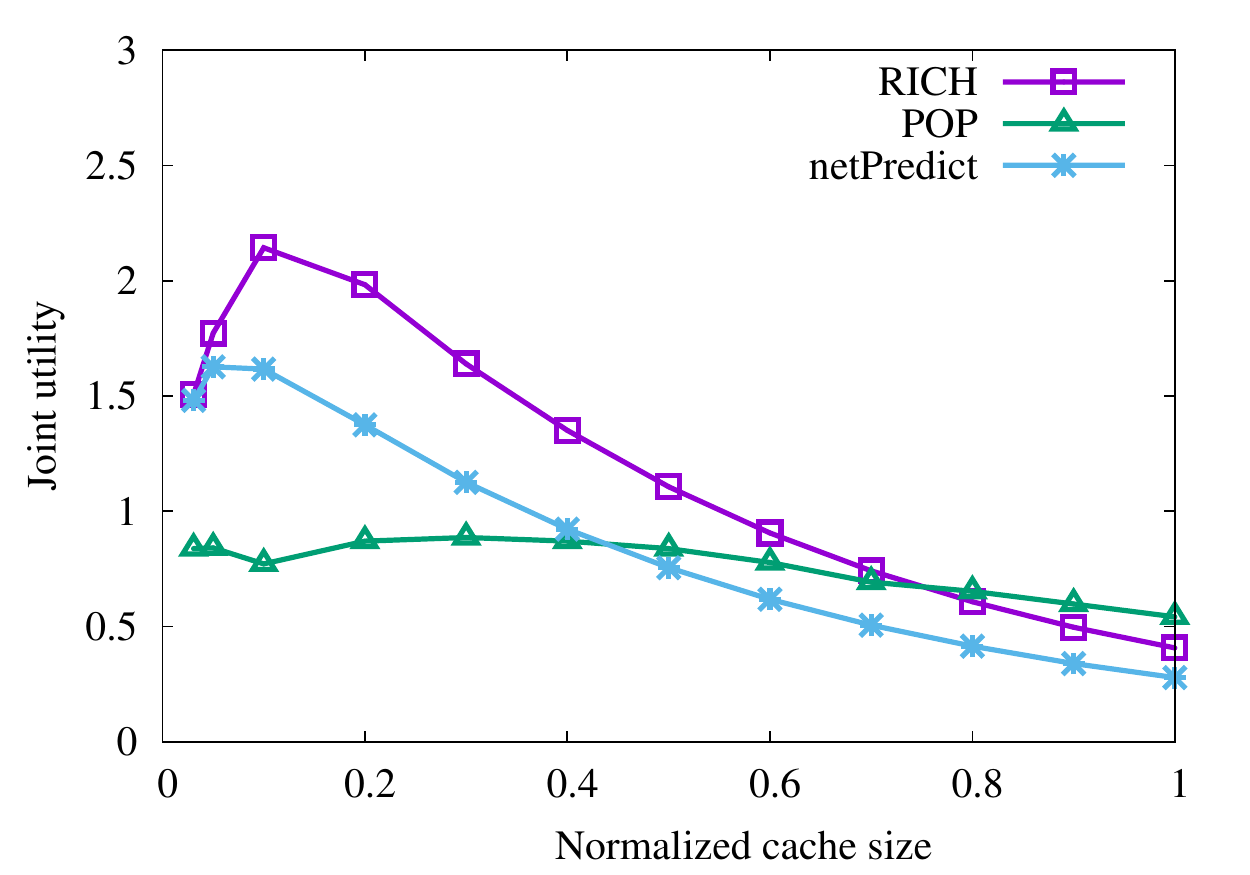}	
	\caption{Urban Bologna scenario: backhaul traffic (left); normalized backhaul overhead (center); joint user/operator utility function (right).} 
	\label{fig:bb}
\end{figure*}

{\bf Edge network performance.} 
We now evaluate the performance of the three policies at the network edge, which directly impacts the content access delay. Fig.~\ref{fig:hit}(left) depicts the {\color{red}cache hit} probability as a function of the normalized cache size $\hat{C}$, for POP, netPredict  and RICH 
prefetching schemes. As expected, a larger cache size improves the performance of all caching schemes, even if beyond $\hat{C}=0.2$ the {\color{red} cache hit} probability for RICH and netPredict becomes constant since under the two policies the cache is never full.
However, RICH  outperforms netPredict and POP up to 33\% and 190\% respectively, for small cache size. 
This is due to the higher effectiveness of the RICH policy, which  tends to store chunks 
only in those ENs from where they can be  downloaded with high probability.
For very large cache size ($\hat{C}>0.75$), POP achieves higher {\color{red}cache hit} probability as compared to RICH. This is because the cache is always fully occupied for POP and at $\hat{C}=1$ each cache stores every content in the catalog, hence a hit is always guaranteed. Fig.~\ref{fig:hit}(center) shows the network cache occupancy for the three caching schemes. 
By construction, regardless of the cache size, the cache occupancy is highest for POP.
In comparison with netPredict, the cache occupancy of RICH is slightly higher but the gain in the other performance metrics (e.g., {\color{red} cache hit} probability, cache throughput and backhaul traffic) is significantly larger. 
Finally, Fig.~\ref{fig:hit}(right) shows that RICH achieves a cache throughput of up to 83~Mbps (i.e., around 11.9~Mbps per single EN/cache), while the maximum value for netPredict is 64~Mbps.








%

{\bf Backhaul network performance.} 
 The performance gain provided by RICH over both netPredict and POP can be observed also in terms of backhaul traffic, 
since the higher {\color{red} cache hit} probability of RICH implies a lower probability to access the server and retrieve the content from there. 
Fig.~\ref{fig:bb}(left) shows the overall backhaul traffic due to cache misses. In the best case, RICH reduces such traffic by approximately 57\% and 70\% as compared to netPredict and POP, respectively. 

{\color{red} Fig.~\ref{fig:bb}(center) highlights that RICH incurs lower backhaul overhead as compared to the other policies for large enough  caches. In the best case, RICH  achieves 27\% and 67\% lower overhead than netPredict and POP, respectively. The lower negative values of the backhaul overhead indicate better reuse of the chunks stored in the cache. For small caches, RICH incurs higher backhaul overhead due to higher number of prefetched chunks and lower chunk reuse, but RICH still achieves lower backhaul traffic as compared to the other schemes (see Fig.~\ref{fig:bb}(left)), thanks to the higher hit probability and content reuse.  }



{\bf A joint user/operator view.} 
Here we compare the performance of the caching schemes in terms of joint user/operator utility functions. The user utility is described as an exponential function decreasing with $1-P_{hit}$, 
since larger $P_{hit}$ implies smaller latency to access the contents. The operator utility instead is modeled as an exponential function decreasing with the normalized cache size. 
Following a standard approach, we express the joint utility function as the product of the user and the operator utility, which is depicted in Fig.~\ref{fig:bb}(right). Note that RICH outperforms  netPredict, regardless of the cache size. Compared to POP,   the RICH utility is significantly higher for small cache sizes, while for larger caches RICH allows for a great reduction in cache occupancy.

\begin{figure}[tb!]
	\centering
		\includegraphics[width=0.35\textwidth]{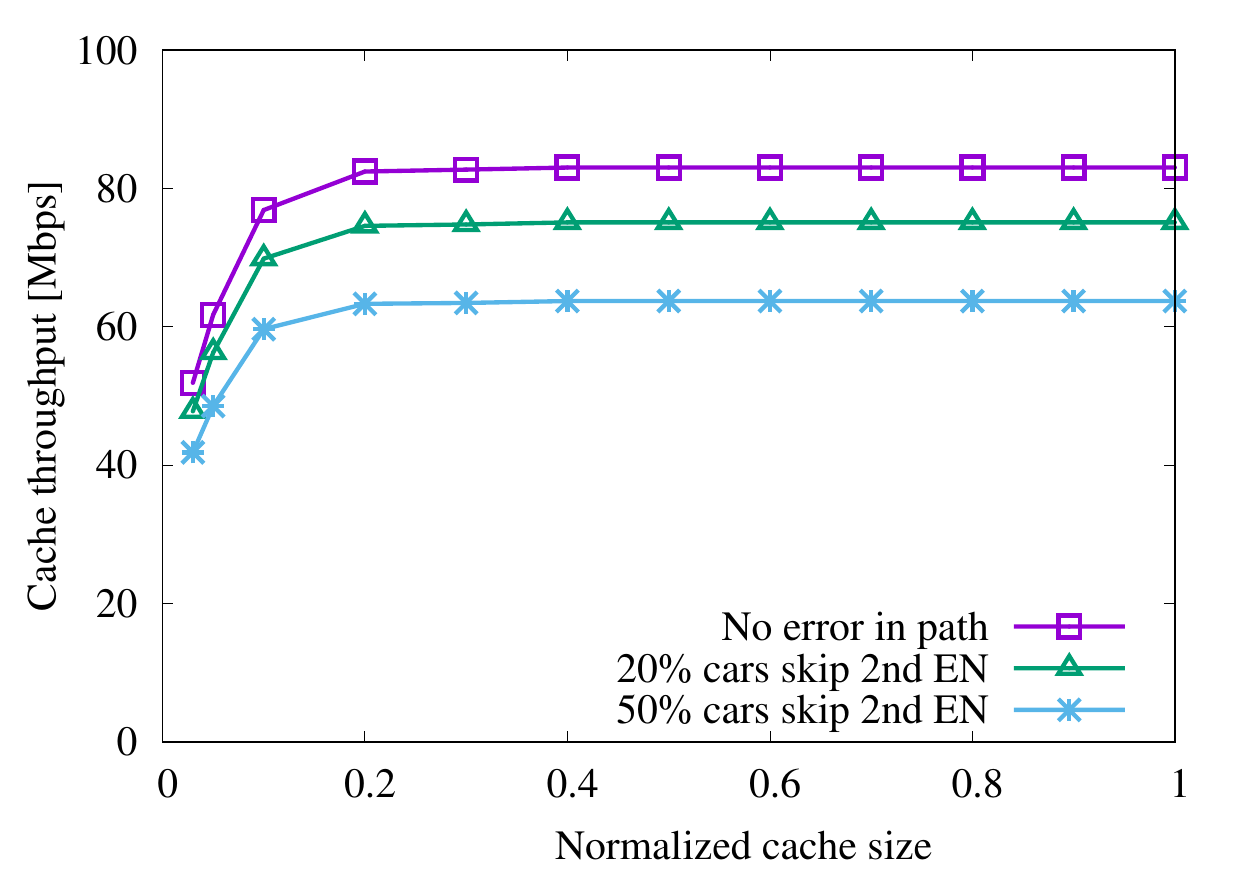}
	\caption{{\color{red}Urban Bologna scenario: effect of error in the car path on the cache throughput.}} 
	\label{fig:cacheTP_errorPath}
\end{figure}

{\color{red}
\subsection{Errors in the knowledge of car mobility}

To assess the robustness of RICH, we evaluate two kinds of error in the system knowledge of the car mobility. The former affects the dwell time under an EN, and, thus,  the actual number of downloaded chunks. The latter affects the knowledge of the  sequence of traversed ENs. 

{\bf Errors in the dwell time.}
We add a random error $\epsilon$  to the dwell time experienced by a car.
Let $W$ be the observed dwell time of a car under a given EN, and let $w_{\min}$ be the  minimum observed dwell time under the same EN, based on past statistics.
We set the actual dwell time $W'$ of the car as:
\[
W'=\max\{w_{\min},W+\epsilon\}
\]
where $\epsilon$ is Gaussian distributed with average $\mu$ and standard deviation $\sigma$. When $\sigma=0$~s, all the dwell times under an EN are deterministically  shifted by  $\mu$.
When  $\epsilon>0$, the car is slowed down,  while, for $\epsilon<0$, the car accelerates with respect to its original speed.

Fig.~\ref{fig:errorDT}(top) shows the cache throughput for different values of $\mu$ and for $\hat{C}=0.4$ in the urban Bologna scenario, where the average dwell time across all the ENs is 39~s.
For $\mu < -50$~s, most of the cars spend the minimum time under the coverage of the ENs ($W'\approx w_{\min}$), thus the cache throughput becomes very low. The cache throughput increases with $\mu$ as the cars get more time to download the chunks available in the cache. For $\mu > 20$~s, instead, the high coverage time  does not have a significant effect on the cache hit events. Indeed, all the stored chunks have already been downloaded and the extra coverage time cannot be exploited, thus the cache throughput decreases. 

Fig.~\ref{fig:errorDT}(bottom) shows the cache hit probability for different values of $\mu$. For $\mu < 0$~s, cars spend less time under coverage, thus both caches hits and misses decrease. However, the cache hit probability remains high due to the large amount of data downloaded by cars under their first EN. For $\mu > 0$~s, the cache hit probability decreases rapidly by increasing $\mu$, due to the higher number of cache miss when the cars experience a coverage time longer than expected. The  hit probability is almost unchanged for $\sigma=0$ and $10$~s, while it  decreases at most by  30\% for $\sigma=60$~s, which is the worst case scenario.

Figs.~\ref{fig:errorDT2}(top) and \ref{fig:errorDT2}(bottom) depict 
the backhaul traffic and the normalized backhaul overhead. Both metrics are low when $\mu < 0$~s, as cars spend less time under coverage, hence the number of cache miss is small and few  requests reach the Data Store. When $\mu > 0$~s, both metrics increase due to the significant increase in the number of cache miss. Notably, due to the large content reuse, the backhaul overhead is negative. 

In summary, only large errors in the dwell time (i.e., large values of $|\mu|$ or  $\sigma$) have an evident impact on the performance, as they make the past statistics nearly useless. Otherwise, the performance is just slightly affected by errors,  confirming the robustness of the proposed approach.

\begin{figure}[tb!]
	\centering
		\includegraphics[width=0.35\textwidth]{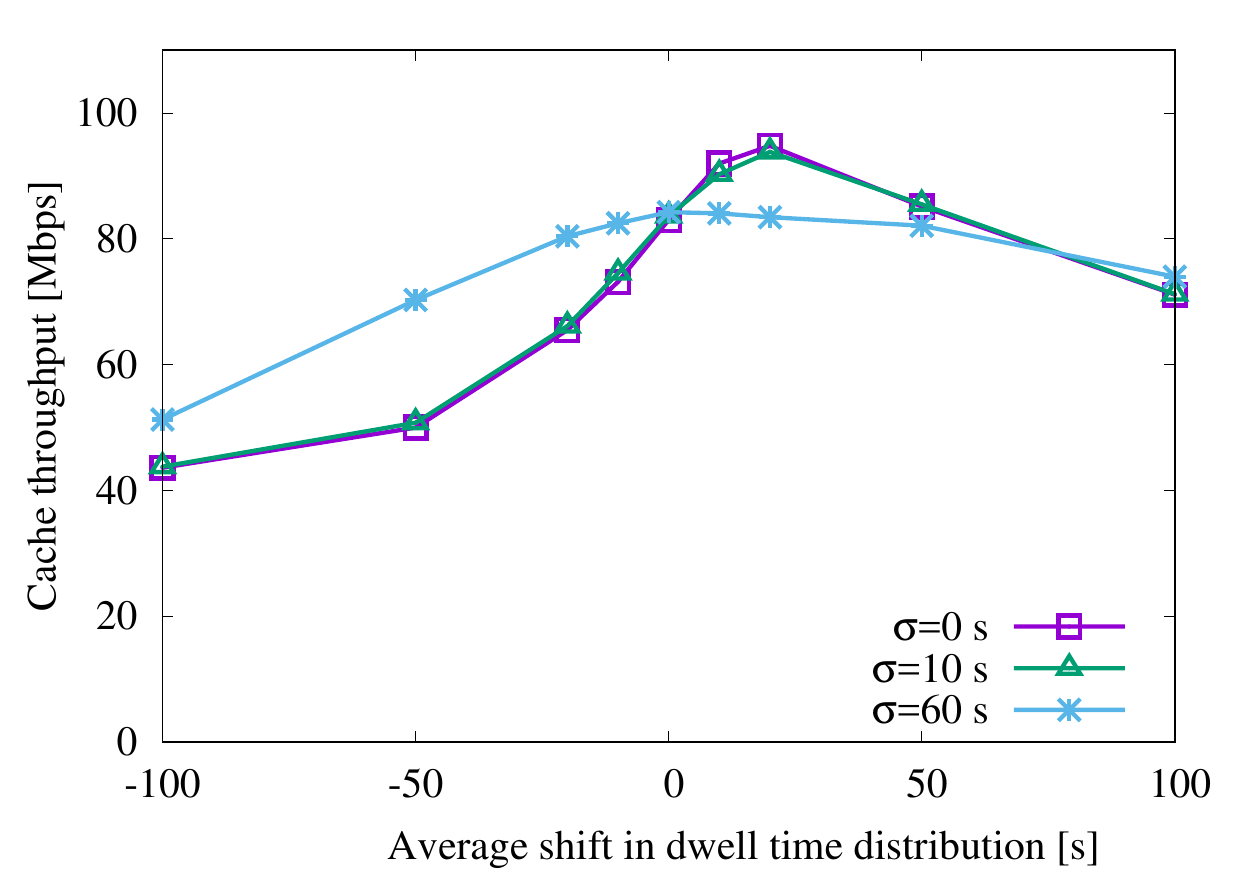}
		\includegraphics[width=0.35\textwidth]{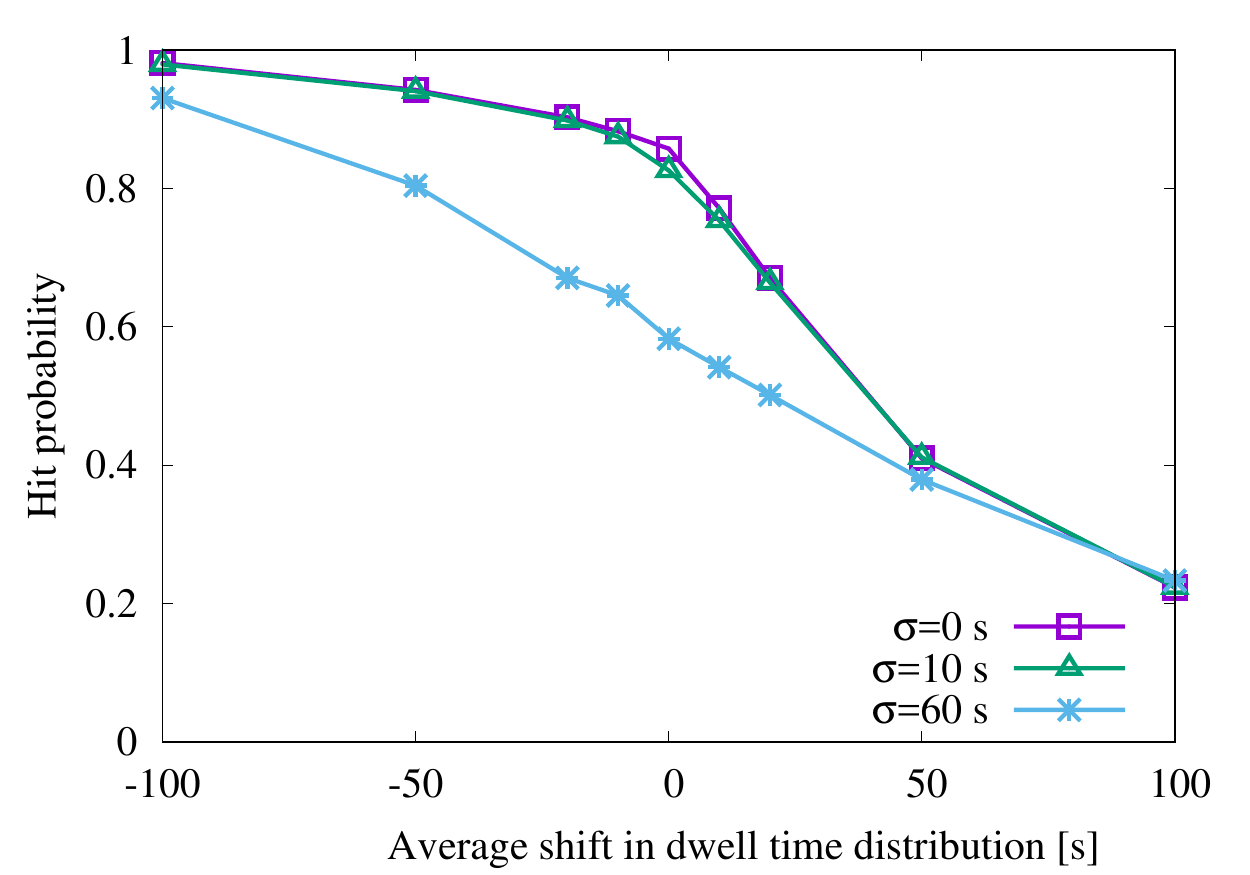}	
	\caption{Urban Bologna scenario: effect of random error $\epsilon$ in dwell time on the cache throughput (top) and on the cache hit probability (bottom). }
	\label{fig:errorDT}
\end{figure}



\begin{figure}[tb!]
	\centering
	\includegraphics[width=0.35\textwidth]{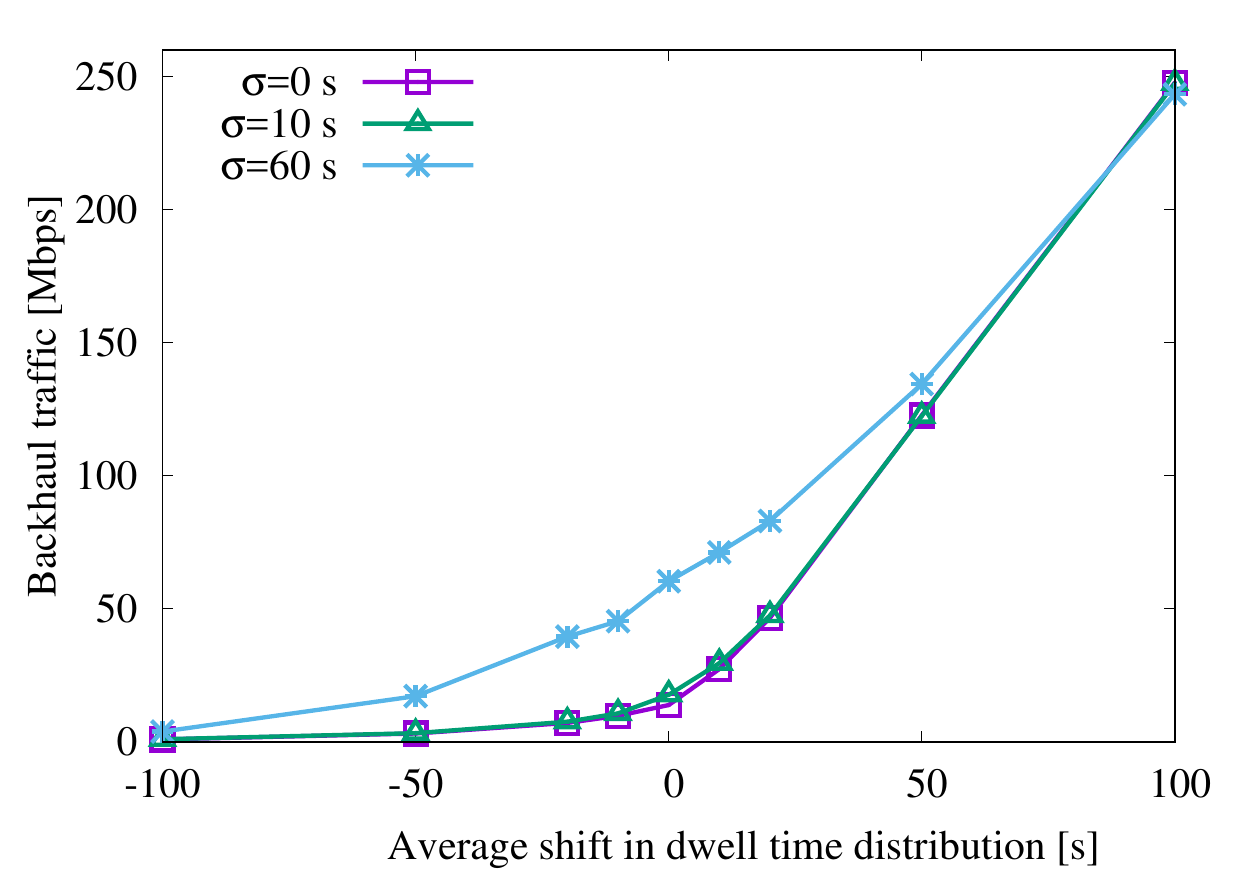}
		\includegraphics[width=0.35\textwidth]{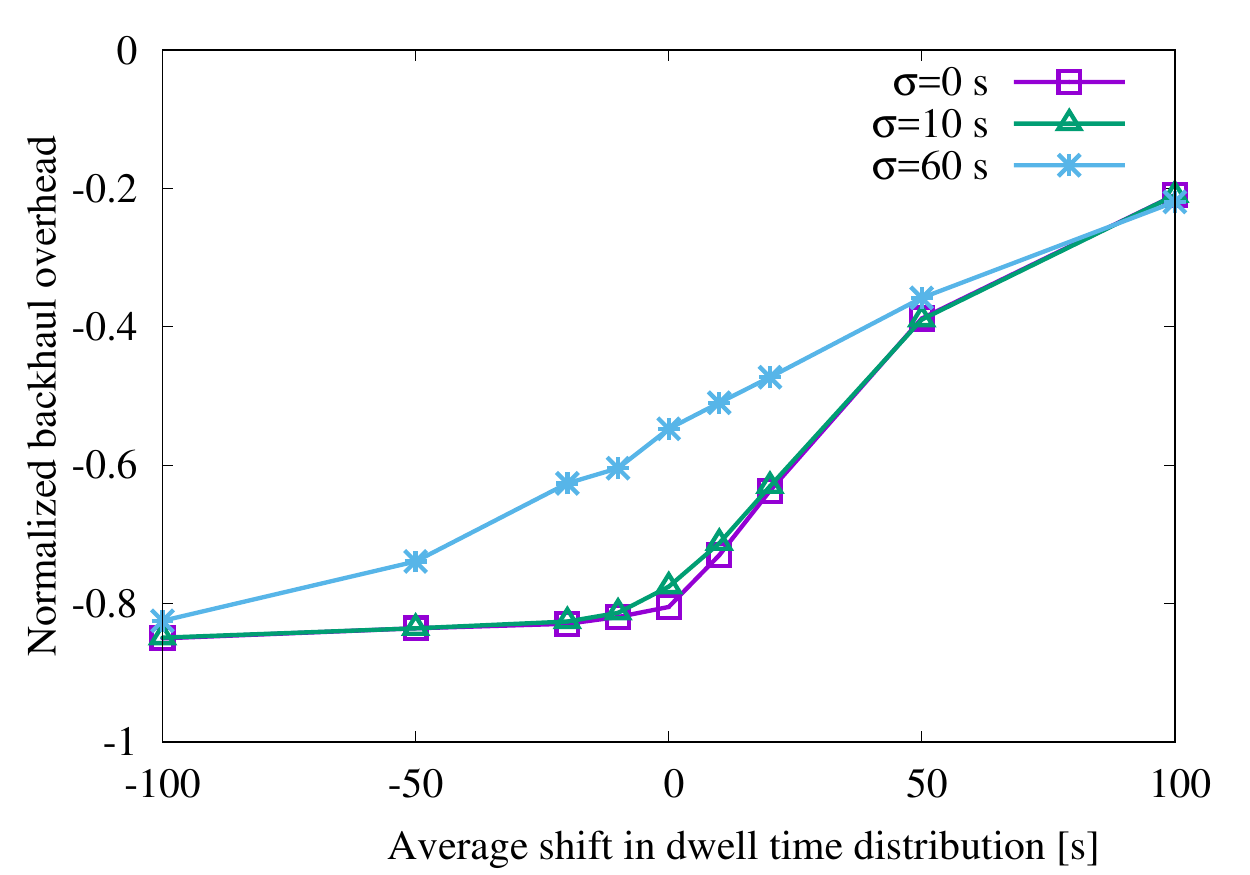}
	\caption{Urban Bologna scenario: effect of random error $\epsilon$ in the dwell time on the backhaul traffic (top) and on the normalized backhaul overhead (bottom). } 
	\label{fig:errorDT2}
\end{figure}



{\bf Errors in the car path.} In the considered scenario, RICH relies on the information on just two subsequent ENs traversed by a car along its path. Additionally, it runs when a car enters the coverage area of the first EN, thus guaranteeing some intrinsic level of robustness. Nevertheless, a car could change its actual trajectory by completely skipping the second EN along its expected path -- an event that could arbitrarily affect the cache hit. 

To understand the effect of errors in the cars path, we selected at random a set of 20\% and 50\% of  cars that would skip the second EN along their path, in the urban Bologna scenario. Fig.~\ref{fig:cacheTP_errorPath} shows the cache throughput (averaged over several simulation runs) versus the cache size. The cache throughput decreases by approximately 10\% and 23\% with respect to the case where cars follow the original path, confirming the robustness of the proposed approach.}

\subsection{{\color{red}More detailed knowledge on car mobility}}
We now analyze the benefit that a more detailed information on the car mobility could bring.
To this end, we classify the cars in two categories: slow and fast. 
Slow cars are those  dwelling under the coverage area of an EN longer than 10\,s; all others are classified as fast cars. Table~\ref{tab:coeffvar} reports the coefficient of variation of the resulting dwell time $\widehat{X}_i$, with and without such classification. As expected, for all ENs the variance of the dwell time for slow and fast cars is smaller, highlighting the gain brought by  the additional information. 

\begin{figure}[tb!]
	\centering
		\includegraphics[width=0.35\textwidth]{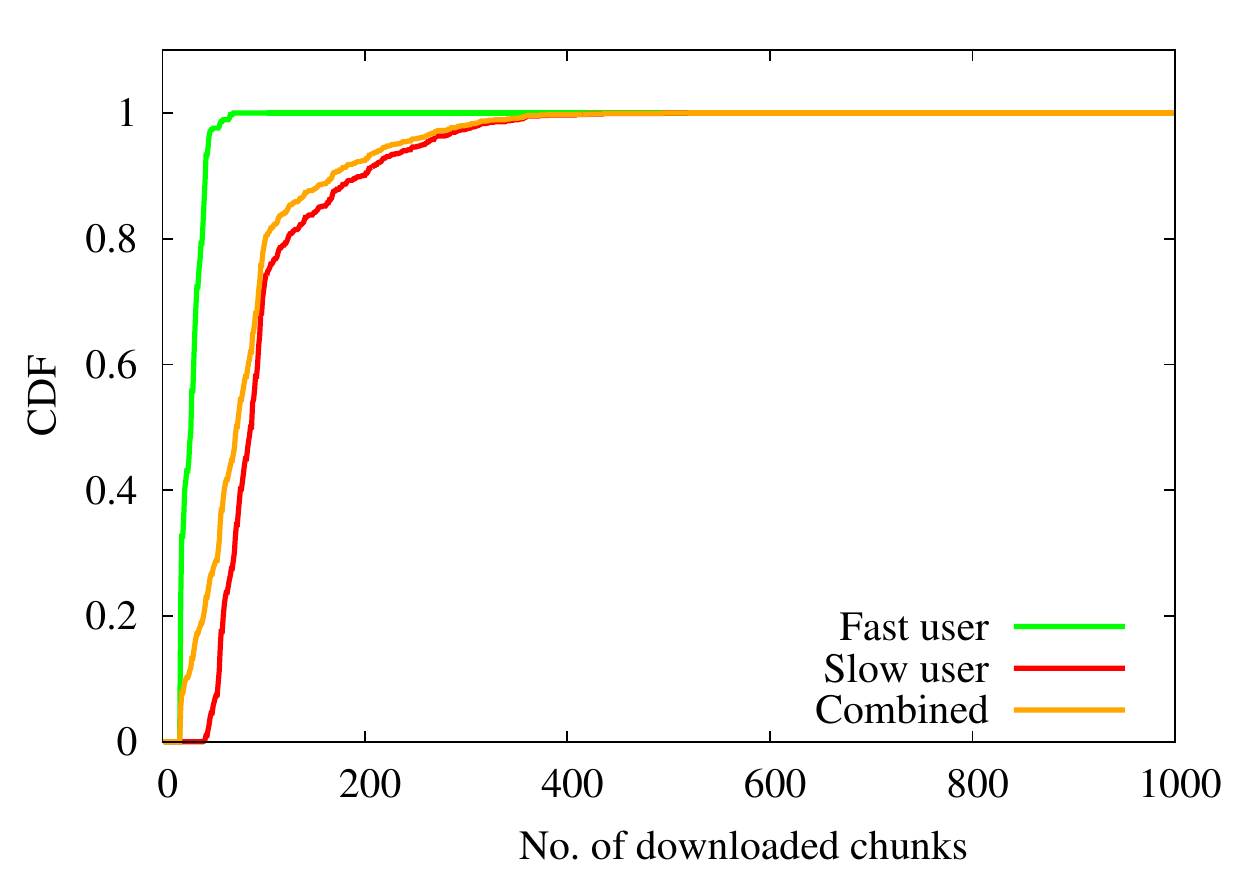}
		\includegraphics[width=0.35\textwidth]{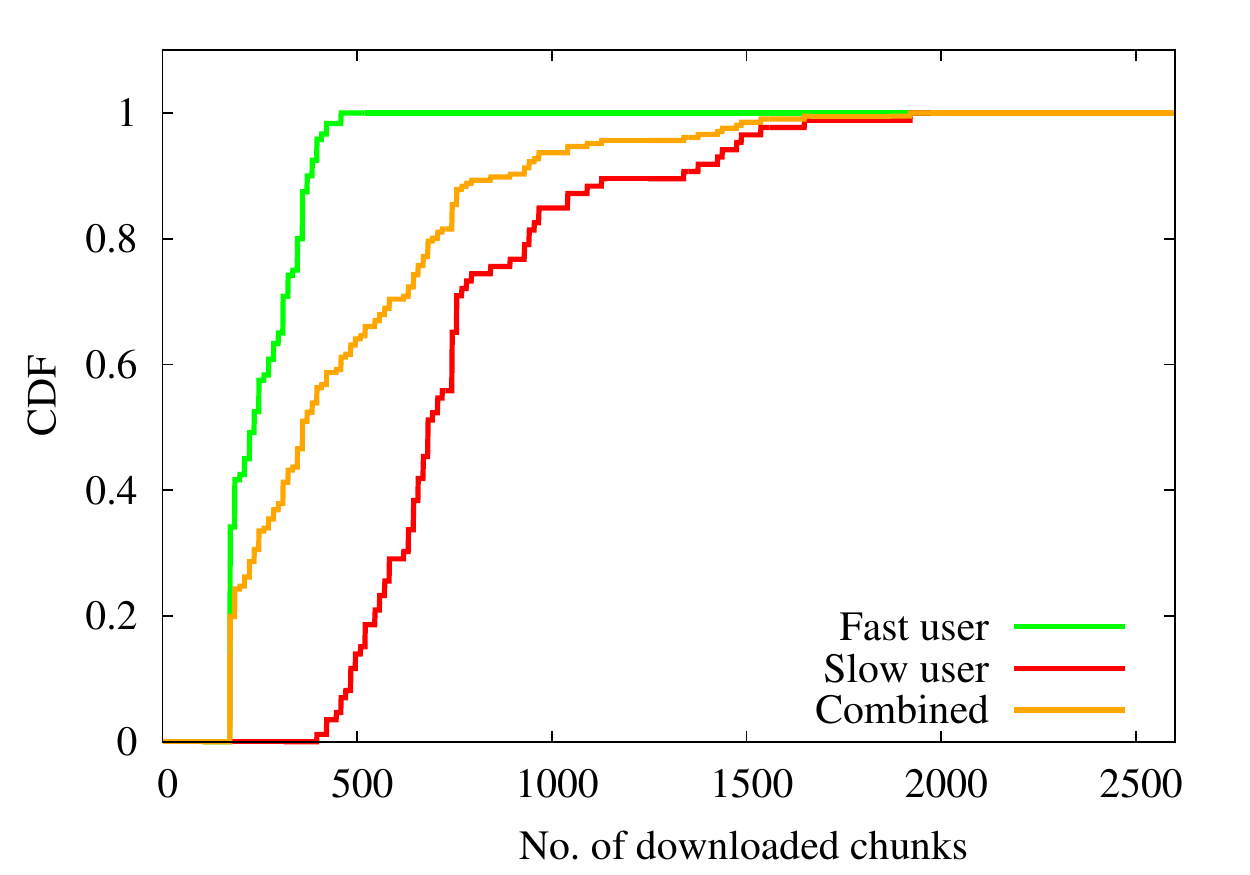}
	\caption{Cumulative density function of the number of downloaded chunks  measured in the trace, based on the slow/fast classification at EN A (top) and at EN H (bottom).}\label{fig:rsuA}
\end{figure}

Figs.~\ref{fig:rsuA}(top) and \ref{fig:rsuA}(bottom) depict the cumulative distribution
corresponding to $\hat{X}_A$ for EN A and $\hat{X}_H$ for EN H, respectively,  in the Bologna urban area, for  cache size equal to 2600 chunks. For EN A,
Fig.~\ref{fig:rsuA}(top) shows that  slow  and fast cars can download, on average, around 100 chunks and  30 chunks, respectively. 
Observing the
average number of cars in Table~\ref{tab:no-users}, EN A is
expected to be located at a very congested intersection. Indeed,  only
few, slow cars experience a large enough dwell time and a small
radio congestion that allow them to download all the 2600 chunks
stored in the cache. EN H, instead,  appears to be located at a much
less congested intersection, according to the values in Table~\ref{tab:no-users}. This
implies a lower radio congestion, hence, a higher download
capability, as shown in Fig.~\ref{fig:rsuA}(bottom). Indeed, under EN H, fast users download at least 170 chunks, 
with an average around
250, whereas slow users download
around 1000 chunks. 
For brevity, we omit the distribution of the number of chunks downloaded from the other ENs; the number of fast and slow cars per EN is reported in Table~\ref{tab:fast-slow}.

Using  the dwell-time distribution conditioned on the fast/slow cars,  we can obtain better results  for both  RICH and netPredict. 
{\color{blue}
The  cache throughput of RICH now reaches about 91~Mbps (10\% increase), while netPredict achieves 83~Mbps (30\% increase). Similarly, the backhaul traffic is further reduced to 6~Mbps (80\% decrease)  for RICH, while it remains unchanged for netPredict. The performance of POP, instead, is unaffected as POP is mobility agnostic. }

\begin{table*}[tb!]
\centering
\caption{Number of fast and slow cars at each EN in the urban Bologna scenario\label{tab:fast-slow}}
\resizebox{0.9\linewidth}{!}{%
\begin{tabular}{cccccccccccccccc}
\toprule
\multicolumn{2} {c}{\textbf{EN A}} & \multicolumn{2} {c}{\textbf{EN B}} & \multicolumn{2} {c}{\textbf{EN C}} & \multicolumn{2} {c}{\textbf{EN D}} & \multicolumn{2} {c}{\textbf{EN E}} & \multicolumn{2} {c}{\textbf{EN F}} & \multicolumn{2} {c}{\textbf{EN G}} & \multicolumn{2} {c}{\textbf{EN H}} \\ \cmidrule(lr){1-2} \cmidrule(lr){3-4} \cmidrule(lr){5-6} \cmidrule(lr){7-8} \cmidrule(lr){9-10} \cmidrule(lr){11-12} \cmidrule(lr){13-14} \cmidrule(lr){15-16} \vspace*{1mm}
Fast & Slow & Fast & Slow & Fast & Slow & Fast & Slow & Fast & Slow  & Fast & Slow & Fast & Slow & Fast & Slow\\ 
376&1204&948&562&182&345&215&167&0&337&1456&0&68&93&120&86 \\ \bottomrule
\end{tabular}%
} 
\end{table*}

\begin{table}[tb!]
\centering
\caption{Coefficient of variation of $\widehat{X}_i$ at each EN $i$.\label{tab:coeffvar}}
\begin{tabular}{cccc}
\toprule
\multirow{2}{*}{\textbf{EN}} & \multicolumn{3} {c}{\textsc{Coefficient of variation}}\\ \cmidrule(lr){2-4}
&\textbf{Fast} & \textbf{Slow} & \textbf{Combined} \\\midrule
A& 0.36& 0.61& 0.75\\
B& 0.31& 0.45& 0.80\\
C& 0.43& 0.32& 0.59\\
D& 0.30& 0.40& 0.72\\
E& --& 	0.32& 0.32\\
F& 0.04& --& 0.04\\
G& 0.38& 0.17& 0.44\\
H& 0.33& 0.40& 0.70\\ \bottomrule
\end{tabular}%
\end{table}

\section{Related work \label{sec:related}}


Several works have appeared on content caching at the edge of the  cellular network. In particular, \cite{geocaching} 
proposes an optimal probabilistic
content placement policy  that maximizes the total {\color{red} cache hit}
probability for random network topologies, based on  content
popularity. Thus, unlike our work, the policy in \cite{geocaching}  is  oblivious to the actual mobility
pattern of the users. A hybrid scenario, comprising MANET and
cellular networks, is studied in \cite{cachinghybrid}, where each node estimates  the content popularity and
caches a content based on that. Importantly, this scheme can
be considered as a distributed implementation of the POP policy that we use 
as benchmark in our work.

Regarding cellular backhaul networks,
\cite{woo2013comparison} investigates the effect of different criteria
to identify a web content  when accessing the caching system.
The main idea is to avoid different application-level identifiers for a content, so as to prevent   content items from being duplicated in the caches. We remark that, in our study, we consider the streaming
of content through a chunk-based approach and we assume that
each chunk and each content are univocally identified.

Several works have employed cooperation among caching nodes to improve performance in heterogeneous cellular networks. As an example, \cite{ma2016cooperation} introduces a hierarchical caching architecture where caches are deployed in both small-cell and macro-cell base stations, and both can satisfy a user content request. 
Similarly, the work in~\cite{chen2017cooperative} considers a cluster of small-cell base stations as one cache entity. The base stations adopt both cooperative caching techniques and cooperative transmissions.
Differently from the above studies, our approach is based on a centralized prefetching scheme orchestrating the caches deployed at the different ENs.


Few works have investigated caching schemes specifically taking into account 
user mobility. {\color{blue} \cite{mobee} proposes a mobility-aware cache placement strategy for Cloud Radio Access Networks.  The authors leverage on the user-mobility pattern to analytically estimate the content-request rate in different cells and minimize the network energy consumption. This work differs from ours  as our goal is to optimize the cache hit probability. \cite{2018ozfatura} proposes a mobility and popularity-aware caching scheme for a heterogeneous cellular network, comprising macro-cell and small-cell base stations, where only the latter are capable to cache contents. The goal of the caching policy is to minimize the data downloaded from the macro-cell base station. 
More relevant to our study is \cite{edgebuffer}, which considers the  MobilityFirst network architecture
introduced in~\cite{mobilityfirst}, and proposes an approach, called netPredict,  
aiming to support smooth mobile content delivery. 
In MobilityFirst, a global identifier is associated with each user, so that the user mobility is recorded at each {\color{blue} network} node. Each of these
nodes is equipped with two distinct buffers, similarly to our architecture in Fig.~\ref{fig:architecture}. The first one caches the
most popular content items, exactly as the POP policy considered in our work.
The second buffer is devoted to store the content based on a
prefetching policy leveraging on the predicted sequence of nodes
traversed by each particular user, and the average 
dwell time and  available bandwidth under each node. 
A similar prefetching policy is
proposed 
in~\cite{dandapat2013} for a cellular network scenario. 
Similarly to our work, the content is delivered to users by  base
stations using a chunk-based approach.  The specific mobility of each
user is used to identify the chunks to prefetch 
along the user path. Unlike our study, however, both~\cite{edgebuffer}
and \cite{dandapat2013} assume that the caching policy knows or predicts
the spatial and temporal trajectory of each user, in order to estimate
the time intervals in which the user will be covered by each base
station. Our approach instead requires only the knowledge of the distribution of
the dwell times under each edge node at aggregate level. This
distribution can be estimated locally by each EN and does {\em not} require  the precise knowledge of the car trajectory: only the sequence of ENs is
needed. This simplifies the prediction process and  provides better user privacy.

Another  body of works relevant to our study is based on the application of the Information-Centric Networking (ICN) paradigm to vehicular networks. In particular, \cite{nextgenCCVN} proposes an architecture for content-centric vehicular networks, which is compatible with our proposed prefetching approach. 
{\color{blue}\cite{khan2016saving} presents a socially-aware vehicular information-centric system, which leverages on caching, computing, and communication capabilities of smart vehicles for facilitating content availability to mobile users.} 
Although in standard ICN-based architectures network nodes reactively store contents during delivery to the user, similarly to our approach \cite{grewe2016perceive} and~\cite{mauri2017optimal} propose to prefetch contents in ICN nodes.  
The study in~\cite{mauri2017optimal} formulates the problem of optimally placing content chunks in the ICN-based network nodes as an integer linear programming optimization problem, maximizing the content retrieval probability. Moreover, forward error correction coding is adopted to reconstruct the whole content if enough chunks have been received, independently from their order. The work in~\cite{mauri2017optimal} is thus based on a single content retrieval and not on content streaming as in the RICH case. 
{\color{blue} Interestingly, the authors in \cite{mobility-aware_caching_ccwn} advocate that it is critical to consider user mobility information for caching design in content-centric wireless networks. They assume that the user dwell time is estimated based on the available data, and optimize the cache failure probability by solving a convex optimization problem. However, the details of the estimator for the dwell time are not provided. 
Similarly, \cite{Cooperative_caching_MEC} proposes a mobility-aware cooperative caching scheme for content-centric 5G networks, where contents can be stored at the network edge as well as in the vehicular cloud. Furthermore, MEC servers are leveraged to compress the contents thereby enhancing the storage capability of the edge nodes. The considered scenario is similar to that of \cite{grewe2016perceive} where vehicles are assumed to be moving at a constant speed, thus, the approach is not applicable to our urban scenario.}

Finally, a preliminary version of this work was included in our conference paper \cite{mahmood2016}, where the basic ideas of our caching framework were sketched.  }

\section{Conclusions\label{sec:conclusions}}

In this paper we study how to efficiently provide
connected cars with streaming data as they drive along a road covered
by wireless Edge Nodes (ENs). Our RICH prefetching policy determines the content chunks to store in the ENs caches, based on the past statistics of the achievable data rates and of the dwell time experienced by the cars under the coverage. RICH requires only to know the sequence of ENs traversed by a car, without any detailed information on its actual trajectory and real-time traffic conditions. 

Throughout extensive trace-driven simulations, our scheme was shown to improve the cache throughput and to reduce the backhaul traffic, with beneficial effects for both the users and the network operators. Furthermore, RICH was shown to be robust to possible errors in the knowledge of the cars path and dwell time.


\section{Acknowledgements}
This work has been partially funded by the EC H2020 5G-Transformer Project (grant no. 761536), the EC H2020 5G-EVE Project and the EC H2020 HIGHTS project (grant no.~636537). 
EURECOM acknowledges the support of its industrial members, namely, BMW Group, IABG, Monaco Telecom, Orange, SAP, ST Microelectronics, and Symantec.


\bibliographystyle{IEEEtran}
\bibliography{biblio}

\appendices
\section{Proof of Corollary 1}\label{sec:app}
\IEEEproof
When $X_{i}$'s are i.i.d., 
\[ 
\phi_i(k)=\sum_{n=1}^{k-1}\PP\left(X\geq k-n \right)\PP\left (Y_{i-1}=n\right ). 
\] 
Thus, (\ref{eq:phi}) can be rewritten as:
\begin{eqnarray}
\phi_i(k) \hspace{-6mm} &\mathord{=}& \hspace{-2mm}\sum_{n=1}^{k-1} \PP \left (X \geq k-n \right) \sum_{t=1}^{n}
\PP \left (X=t | Y_{i-2}=n-t \right) \cdot \nonumber\\
        \hspace{-6mm}& &\hspace{-2mm} \PP \left ( Y_{i-2}=n-t)\right ) \nonumber\\
\hspace{-6mm}&\stackrel{z=n-t}{=}& \hspace{-2mm}\sum_{z=1}^{k-1} \PP\left (X\geq (k-t) -z \right) \cdot\nonumber\\
       \hspace{-6mm}& &\hspace{-2mm}\sum_{t=1}^{k-1} \PP \left (X=t | Y_{i-2}=z \right) \PP\left ( Y_{i-2}=z \right) \nonumber\\
\hspace{-6mm}&=& \hspace{-2mm} \sum_{t=1}^{k-1} \PP \left (X=t \right) \cdot \nonumber\\
        \hspace{-6mm}& &\hspace{-2mm} \sum_{z=1}^{k-1} \PP\left ( X\geq (k-t) -z \right) \PP \left ( Y_{i-2}=z \right ) \nonumber\\
\hspace{-6mm}&=& \hspace{-2mm}( f_X \ast\phi_{i-1})(k) \,.
\label{eq:iid-conv}
\end{eqnarray}
\endIEEEproof

\end{sloppypar}
\end{document}